\documentclass[11pt,letterpaper]{article}
\usepackage[normalem]{ulem}
\usepackage{jheppub}
\usepackage[english]{babel}
\usepackage{amssymb,amsmath}
\usepackage{mathtools}
\usepackage{mathrsfs}
\usepackage{bbold}
\usepackage{hyperref}
\usepackage{empheq}
\usepackage{subfigure}
\usepackage{xfrac} 
\usepackage{graphics,color}
\usepackage{slashed}
\usepackage{graphicx}
\usepackage{braket}
\usepackage{float}
\usepackage{tcolorbox}
\usepackage{booktabs}
\usepackage[normalem]{ulem}

\newcommand{\Mpl}{M_{\rm Pl}}
\newcommand{\be}{\begin{equation}}
\newcommand{\ee}{\end{equation}}
\newcommand{\bea}{\begin{eqnarray}}
\newcommand{\eea}{\end{eqnarray}}
\newcommand{\nn}{\nonumber}

\usepackage{tikz}
\usetikzlibrary{arrows,shapes}
\usetikzlibrary{trees}
\usetikzlibrary{matrix,arrows} 				
\usetikzlibrary{positioning}				
\usetikzlibrary{calc,through}				
\usetikzlibrary{decorations.pathreplacing}  
\usepackage{pgffor}							

\usetikzlibrary{decorations.pathmorphing}	
\usetikzlibrary{decorations.markings}
\tikzset{
    vector/.style={decorate, decoration={snake}, draw},
	provector/.style={decorate, decoration={snake,amplitude=2.5pt}, draw},
	antivector/.style={decorate, decoration={snake,amplitude=-2.5pt}, draw},
	graviton/.style={decorate, decoration={snake,amplitude=1.5pt}, draw},
    fermion/.style={draw=black, postaction={decorate},
        decoration={markings,mark=at position .55 with {\arrow[draw=black]{>}}}},
    fermionbar/.style={draw=black, postaction={decorate},
        decoration={markings,mark=at position .55 with {\arrow[draw=black]{<}}}},
    fermionnoarrow/.style={draw=black},
    gluon/.style={decorate, draw=black,
        decoration={coil,amplitude=4pt, segment length=5pt}},
    scalar/.style={dashed,draw=black, postaction={decorate},
        decoration={markings,mark=at position .55 with {\arrow[draw=black]{>}}}},
    scalarbar/.style={dashed,draw=black, postaction={decorate},
        decoration={markings,mark=at position .55 with {\arrow[draw=black]{<}}}},
    scalarnoarrow/.style={dashed,draw=black},
    electron/.style={draw=black, postaction={decorate},
        decoration={markings,mark=at position .55 with {\arrow[draw=black]{>}}}},
	bigvector/.style={decorate, decoration={snake,amplitude=4pt}, draw},
}

\title{Instanton NDA and Applications to Axion Models}
\preprint{}

\author[a]{Csaba Cs\'aki,}
\author[b,c]{Raffaele Tito D'Agnolo,}
\author[d]{Eric Kuflik,}
\author[a]{and Maximilian Ruhdorfer\,}

\emailAdd{csaki@cornell.edu}
\emailAdd{raffaele-tito.dagnolo@ipht.fr}
\emailAdd{eric.kuflik@mail.huji.ac.il}
\emailAdd{m.ruhdorfer@cornell.edu}

\affiliation[a]{Laboratory for Elementary Particle Physics, Cornell University, Ithaca, NY 14853, USA}
\affiliation[b]{Universit\'e Paris-Saclay, CEA, CNRS, Institut de Physique Th\'eorique, 91191, Gif-sur-Yvette, France}
\affiliation[c]{Laboratoire de Physique de l’École Normale Supérieure, ENS, Université PSL, CNRS, Sorbonne Université, Université Paris Cité, F-75005 Paris, France}
\affiliation[d]{Racah Institute of Physics, Hebrew University of Jerusalem, Jerusalem 91904, Israel}

\abstract{We present a simple set of power counting rules which  allows us to easily estimate calculable instanton effects up to ${\cal O}(1)$ factors. We apply the resulting Instanton NDA to examine the effects of small instantons on various axion models. We confirm that mechanisms that increase  the axion mass via small instantons  generically also lead to an enhancement of misaligned instanton contributions to the axion potential, deepening the axion quality problem.  For generic models, new sources of CP violation in the UV must be absent in order to raise the axion mass above the QCD prediction.   
However, we find that $Z_N$ and composite axions are UV-safe against these misalignment effects. Axion GUT models are   also insensitive to UV contributions at the GUT scale, unless a very large number of extra states are introduced below this scale. }

\makeatletter
\@addtoreset{footnote}{section}
\makeatother

\begin{document}

\maketitle	

\section{Introduction}
%
Instantons play a prominent role in many areas of particle physics: they are responsible for the topologically non-trivial $\theta$-vacuum for non-abelian gauge theories, when the gauge coupling is small~\cite{Belavin:1975fg,tHooft:1976snw,Jackiw:1976pf,Callan:1976je,tHooft:1976rip}, they are an important source of baryon number violation in the standard model (SM)~\cite{tHooft:1976rip,Ringwald:1989ee,Espinosa:1989qn,McLerran:1989ab}, and are also responsible for numerous non-perturbative effects in asymptotically free supersymmetric (SUSY) gauge theories (such as the Affleck-Dine-Seiberg superpotential for certain numbers of flavors and colors~\cite{Affleck:1983mk,Davis:1983mz,Cordes:1985um} or the structure of the Seiberg-Witten moduli space for theories with Coulomb branches~\cite{Seiberg:1994rs,Seiberg:1994aj}). While the mass of the QCD axion itself is not due to an instanton effect~\cite{Witten:1978bc,Veneziano:1979ec,Witten:1979vv}, the axion potential remains sensitive to corrections from small size instantons (``small instantons")~\cite{Holdom:1982ex,Holdom:1985vx,Flynn:1987rs,Gherghetta:2020keg,Agrawal:2017ksf,Csaki:2019vte,Dine:1986bg,Choi:1998ep,Rubakov:1997vp}. Performing an instanton calculation is usually quite tedious, and involves the use of the explicit expressions of the fermionic zero modes in an instanton background, as well as integrating over the zero and non-zero mode fluctuations. It is natural to ask whether there are some simple power counting rules that could reproduce the bulk of the effects of the full instanton calculation (up to possible $\mathcal{O}(1)$ factors). In this paper we explicitly present the rules of an ``Instanton NDA", which should be viewed as the analog of the usual Naive Dimensional Analysis (NDA) for ordinary Feynman diagrams~\cite{Manohar:1983md}. Our rules will correctly account for the $4\pi$ factors appearing due to insertions of fermion zero modes or loop integrals, and are typically accurate up to factors of few. 

Our main applications of Instanton NDA will be to carefully examine the effects of small instantons on the axion potential and on the neutron EDM in various axion models. The QCD axion is introduced to solve the strong CP problem, and it generally suffers from the so-called axion quality problem: the Peccei-Quinn symmetry (whose breaking produces the axion) has to be very close to an exact symmetry, because even very high-dimensional operators, suppressed by the Planck scale, can misalign the axion potential and reintroduce a large correction to $\bar\theta$, the parameter entering the neutron EDM.

A less  appreciated aspect of the axion quality problem is that there is another potential source of misalignment for $\bar \theta$. New CP violating sources at high energies can enter the axion potential via small instanton effects. This is a perfect arena to demonstrate the power of Instaton NDA. We show that by inspecting just a few diagrams one can easily decide which models suffer from this source of misalignment and which ones are UV-safe.

One class of models that is sensitive to these corrections are those where the axion mass is enhanced by slowing down the running of the QCD gauge coupling in the UV, by adding more matter. The enhanced gauge couplings in the UV enhance small instanton corrections, and, depending on the details of the model, they could even dominate over the usual IR contributions due to confinement in QCD. We show (following \cite{Bedi:2022qrd}) that this enhancement generically also increases the misalignment effects from small instantons, if new sources of CP violation are present in the UV completion of the theory. This makes it very difficult to have a successful  solution to the strong CP problem. Hence generic models of enhanced axion mass should also explain why new sources of CP violation are absent in the UV.   

We find, however, that two mechanisms~\cite{Hook:2018jle,Contino:2021ayn}, recently introduced to solve the axion quality problem, are generically quite safe against these new misalignment effects originating from small instantons interacting with new UV sources of CP violation. Misalignment effects in composite axion models~\cite{Contino:2021ayn} are generically negligibly small, and they also remain very small in the $Z_N$ axion model of Hook~\cite{Hook:2018jle}, as long as one doesn't introduce too much explicit breaking of the $Z_N$ symmetry. The presence of such explicit breaking terms can, however, potentially spoil the model, placing a lower bound on the scale of UV completion that could be much higher than in models without the explicit breaking. Finally we also examine axion GUT models, and find that small instantons are strongly suppressed, unless a very large set of new particles in complete unified multiplets is added below the GUT scale. 

The paper is organized as follows: In Section~\ref{sec:InstantonCalculus} we introduce Instanton NDA and give a few simple examples of its use in the calculation of the axion potential. We also comment on how the estimates can be promoted to a full instanton calculation. In Section~\ref{sec:examples} we apply Instanton NDA to demonstrate how small instantons can give a contribution to the axion potential comparable to confining dynamics in the IR. Section~\ref{sec:misalignment} is dedicated to the study of misaligned contributions to the axion potential from small instantons in theories with CP violating higher-dimensional operators. In Section~\ref{sec:UVSafe} we demonstrate that $Z_N$ axion and composite axion models are safe against misaligned small instanton contributions to their potential. In Section~\ref{sec:GUTs} we finally discuss small instantons in GUT models. We conclude in Section~\ref{sec:conclusions}.
%
\section{Instanton NDA vs. Full Instanton Calculations}\label{sec:InstantonCalculus}
%
The axion $a$ is the Goldstone boson (GB) of a non-linearly realized chiral $U(1)_{\rm PQ}$, the Peccei-Quinn (PQ) symmetry, which has a mixed anomaly with QCD. The anomaly induces a coupling of the axion to the QCD field strength $G_{\mu\nu}$ and its dual $\tilde{G}_{\mu\nu}$
\be \label{eq:axionLag}
\mathcal{L}_a \supset \left( \bar{\theta} + A\frac{a}{f_a}\right) \frac{g^2}{32\pi^2} \sum_{a=1}^8G^a_{\mu\nu}\tilde{G}^{a\,\mu\nu}\,,
\ee
where $g$ is the QCD coupling, $f_a$ the axion decay constant and $A$ a model dependent anomaly coefficient. The PQ symmetry acts on the axion as a continuous shift symmetry: $a/f_a \rightarrow a/f_a + \alpha$. If unbroken, the symmetry makes the angle $\bar{\theta}$ unphysical. However, in the instanton background, the topological charge is non-vanishing $Q = \langle \tfrac{g^2}{32\pi^2} G^a_{\mu\nu}\tilde{G}^{a\,\mu\nu}\rangle_{\rm inst} \in \mathbb{Z}$, and quantized, such that the continuous shift-symmetry is broken to a discrete shift-symmetry of the form $a/f_a \rightarrow a/f_a + 2\pi n/A$, $n\in\mathbb{Z}$. For simplicity we will take $A=1$ in the following.

Since instantons  break the continuous axion shift symmetry explicitly, they also generate a potential for the axion. They give the leading contribution to the potential when the gauge coupling is small ($g^2 \hslash \ll 1$). At low energy, where QCD confines, the axion potential should be computed using the Chiral Lagrangian, which also incorporates additional effects of the strong dynamics besides those from instantons, as discussed in Section~\ref{sec:lowE}. However, in this work we discuss mostly UV effects that can be computed via reliable instanton calculations at small coupling. 

An explicit computation of the potential including all $\mathcal{O}(1)$ factors is typically technically involved and for most purposes it is not necessary to fix all $\mathcal{O}(1)$ coefficients. In such occasions a few simple power counting rules are in general sufficient to identify and estimate instanton contributions to the axion potential. The goal of this section is to provide the reader with the necessary tools to understand instanton effects and perform quick back-of-the-envelope estimates. In Section~\ref{sec:full}, we make our approach more rigorous and explain how a fully accurate instanton computation can be performed.
%
\subsection{Instanton Basics} \label{sec:InstPowerCounting}
%
The QCD or Yang-Mills vacuum is a superposition of degenerate, but topologically distinct, vacua, the so-called ``$n$-vacua'', characterized by (assuming a proper gauge choice) an integer winding number $n$ of the gauge field at infinity (see e.g.~\cite{Shifman:2012zz} for a pedagogical introduction). Instantons are localized and topologically stable gauge field configurations in Euclidean spacetime that interpolate between the $n$-vacua with different winding number. They are solutions of the Euclidean equations of motion and are therefore saddle points of the action. At small gauge coupling one can show that one-instanton solutions (i.e. solutions of the equation of muotion characterized by a topological charge of $Q=\pm 1$) are the dominant saddle point. Correlation functions in the one-instanton background can then be obtained by performing a semi-classical expansion of the Euclidean action around the instanton solution
\be \label{eq:InstantonAction}
S = S_0 + \int d^4 x \sum_i \delta\Phi_i \mathcal{M}_{\Phi_i} \delta \Phi_i = \frac{8\pi^2}{g^2} \pm i \frac{a}{f_a} + (\text{quadratic fluctuations})\,,
\ee
where $S_0 = 8\pi^2/g^2 \pm i a/f_a$ is the classical action in the instanton background and $\delta\Phi_i$ are quantum fluctuations around the classical field values in the instanton background. The axion dependent term in the classical action is the $\tfrac{a}{f_a} G \tilde{G}$ term in Eq.~\eqref{eq:axionLag} evaluated in the instanton ($+$) or anti-instanton ($-$) background, respectively, with $\bar{\theta}$ absorbed in the axion field. The computation of the quantum fluctuations in the instanton background has been performed by 't Hooft in~\cite{tHooft:1976snw}. Performing the calculation one finds that
\begin{itemize}
    \item Instantons are characterized by their spacetime position $x_0$, their size $\rho$ and their orientation within the gauge group. A change in any of these quantities leads to an equally valid instanton solution, i.e. these so-called collective coordinates are flat directions, also called zero-modes, in the path integral and the integral over them has to be separated out
    \be \label{eq:FIDecomp}
        \int \mathcal{DA} = \widetilde{C}_N \left(\frac{8\pi^2}{g^2}\right)^{2N}\int d\kappa \int d^4x_0\int \frac{d\rho}{\rho^5}\,  \left(\rho \mu_0 \right)^{4N}\int\mathcal{D\tilde{A}}\,,
    \ee
    where $\int d\kappa$ and $\int\mathcal{D\tilde{A}}$ denote the integral over the group orientation and the non-zero modes of the gauge field, respectively. The prefactors $\widetilde{C}_N$, $(8\pi^2/g^2)^{2N}$ as well as $\rho^{-5}$ and $\left(\rho \mu_0 \right)^{4N}$ originate from the Jacobian of the transformation to collective coordinates, where we included a regulator mass scale $\mu_0$ which is introduced when regularizing the UV divergences in the integral over non-zero modes. This is most easily done using Pauli-Villars regularization in which case $\mu_0$ is the regulator mass, i.e. the UV cutoff. Note that $g$ in Eq.~\eqref{eq:FIDecomp} is the bare gauge coupling. One expects that two loop effects will introduce a scale dependence for the coupling. In the following we will ignore two-loop effects and always evaluate the coupling at the scale where the integral over the instanton size is dominated.
    \item The path integral over non-zero modes of the gauge and matter fields, in addition to the $(\rho \mu_0)$ dependence from the zero-modes and UV regulator, makes the coupling in Eq.~\eqref{eq:InstantonAction} scale dependent\footnote{Note that in supersymmetric theories the full scale dependence originates from the zero-modes.} and modifies the numeric prefactor $\widetilde{C}_N$, i.e. 
    \be\label{eq:zeroModeEffect}
        (\rho \mu_0 )^{4N} e^{-\frac{8\pi^2}{g^2}} \rightarrow (\rho \mu_0 )^{b_0} e^{-\frac{8\pi^2}{g^2}} = e^{-\frac{8\pi^2}{g^2(1/\rho)}} \,,\qquad \widetilde{C}_N \rightarrow C_N\,,
    \ee
    where we took the tree-level gauge coupling to be defined at $\mu_0$, i.e. $g \equiv g(\mu_0)$, and $b_0$ is the one-loop beta function coefficient. The explicit form of $C_N$ is given in Eq.~\eqref{eq:CN}.
    \item The Dirac operator of fermions in non-trivial representations under the gauge group has $2k$ zero modes in the instanton background. $k$ is the Dynkin index of the representation which we take to be normalized as $k=1/2$ and $k=N$ for the fundamental and adjoint representations of $SU(N)$, respectively. Therefore, after perfoming the gaussian part of the path integral over the other modes, we are still left with the integral over the $2k$ zero modes. This implies that any non-vanishing correlation function needs at least $2k$ insertions of each fermion which transforms non-trivially under the gauge group.
\end{itemize}
Thus any non-vanishing correlation function in the instanton background is schematically of the form
\be\label{eq:tHooftCorrelator}
\mathcal{N}\int\mathcal{DA}\mathcal{D}\psi\, \underbrace{\psi\cdots \psi}_{2k\text{ times}} e^{-S} \sim C_N \left(\frac{8\pi^2}{g^2}\right)^{2N} \int d\kappa \int d^4x_0\int \frac{d\rho}{\rho^5} e^{-\frac{8\pi^2}{g^2(1/\rho)} \pm i \frac{a}{f_a}} \underbrace{\psi^{(0)} \cdots \psi^{(0)}}_{2k\text{ times}}\,,
\ee
where $\psi^{(0)}$ is the zero-mode wavefunction whose explicit form for a fermion in the fundamental representation of $SU(N)$ is shown in Eq.~\eqref{eq:FermZeroMode} and $C_N$ is the constant which we introduced in Eq.~\eqref{eq:zeroModeEffect}. 
Its value for $SU(N)$ is given in Eq.~\eqref{eq:CN}. For asymptotically free theories the integral over the instanton size is dominated by large IR instantons where the coupling gets large and the exponential suppression vanishes. 
This is due to the logarithmic running of the gauge coupling in the exponential instanton factor which is generated after performing the path integral over non-zero modes (see Eq.~\eqref{eq:zeroModeEffect}). The coupling evolution $\tfrac{8\pi^2}{g^2(1/\rho)} = \tfrac{8\pi^2}{g^2(\mu_0)}-b_0 \log \mu_0\rho$, where $\mu_0$ is a reference scale and $b_0$ the one-loop beta function coefficient,
gives an additional factor of $\rho^{b_0}$ in the integrand. For $b_0>0$, as is the case for asymptotically free theories, this makes the integral more IR dominated. However, in this region instantons are no longer the dominant saddle point of the path integral, the integral becomes IR divergent, and the one-instanton calculation does not give reliable results. We will come back to this issue momentarily.

The above discussion implies that the effects of an instanton can be captured by a local fermion operator, the so-called 't Hooft operator. Schematically this operator is of the form
\be
\vcenter{\hbox{\includegraphics[width=0.25\textwidth]{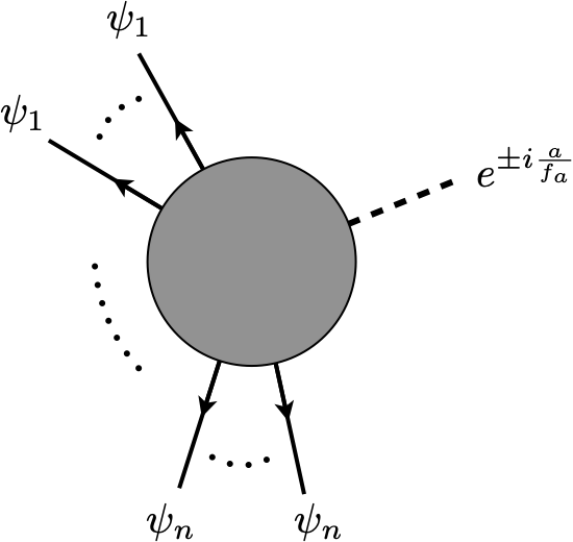}}} \sim C_N \left(\frac{8\pi^2}{g^2(M_{\rm IR})}\right)^{2N} e^{\pm i \frac{a}{f_a}}\frac{e^{-\frac{8\pi^2}{g^2(M_{\rm IR})}}}{M_{\rm IR}^{3(k_1+\ldots +k_n)-4}} \psi_1^{2k_1}\cdots \psi_n^{2k_n}\,,
\ee
and captures the effect of instantons of size $\rho < 1/M_{\rm IR}$ in the UV theory where $M_{\rm IR}$ is an IR cutoff. Note that in order to arrive at the exact expression for the local 't Hooft operator (which can then be used as an ordinary vertex in a Feynman diagram) one has to apply the LSZ reduction formula to the correlation function in Eq.~\eqref{eq:tHooftCorrelator}. 
In the previous expression we also evaluate the gauge coupling in the prefactor at $M_{\rm IR}$ where the integral over the instanton size is dominated. Note that the integral over the instanton orientation within the gauge group ensures gauge invariance and fixes the index contractions of the fermion fields. It is often convenient to trade $e^{-\frac{8\pi^2}{g^2(M)}}$ for the RG invariant scale $\Lambda_G$ which is defined as 
\be
\Lambda^{b_0}_G \equiv M^{b_0}e^{-\frac{8\pi^2}{g^2(M)}}\,, \label{eq:RGI} 
\ee
where the subscript $G$ stands for the gauge group. For asymptotically free theories $\Lambda_G$ is the one-loop definition of the confinement scale.

\subsection{Instanton NDA}\label{sec:NDA}

With this picture in mind we can formulate a set of simple rules to obtain instanton contributions to the axion potential. The axion potential can be obtained from the vacuum-to-vacuum amplitude in the one-(anti)-instanton background, 
which can be obtained by closing the fermion legs of the 't Hooft operator, i.e. we have to soak up the fermion zero modes. Let us also stress at this point that the fermionic legs of the 't Hooft operator stand for zero-mode wavefunctions, i.e. some caution is required when interpreting it as an effective operator. An estimate for 
the contribution to the axion potential can be obtained by following these simple steps:
\begin{enumerate}
    \item Identify the 't Hooft operator. It contains one leg for each fermionic zero mode. The number of zero modes for each fermion is given by $2k$. $k$ is the Dynkin index of the representation under the gauge group, in the convention where $k=1/2$ for the fundamental representation. 
    For example in an $SU(N)$ gauge theory with two flavors of vector like fermions in the fundamental representation $\psi_1,\bar{\psi}_1$ and $\psi_2, \bar{\psi}_2$ the 't~Hooft vertex looks like
\be \label{eq:step1}
\vcenter{\hbox{\includegraphics[width=0.25\textwidth]{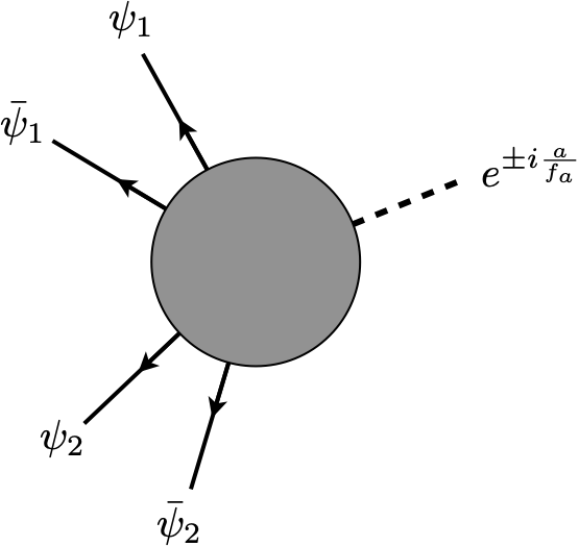}}} 
\ee
    \item Treat the 't Hooft operator as a $2k$-point vertex in a Feynman diagram and  close the fermion legs using any available coupling,  as you would do for an ordinary vacuum-to-vacuum amplitude. Particles with mass $m < 1/\rho$ propagate freely, and can appear in loops, those of mass $m>1/\rho$ should be integrated out, and the resulting effective operators can be used to close fermion legs.
    The 't~Hooft vertex in Eq.~\eqref{eq:step1} can e.g. be closed with Yukawa couplings and a loop of light scalars $\phi$
\be \label{eq:step2}
\vcenter{\hbox{\includegraphics[width=0.7\textwidth]{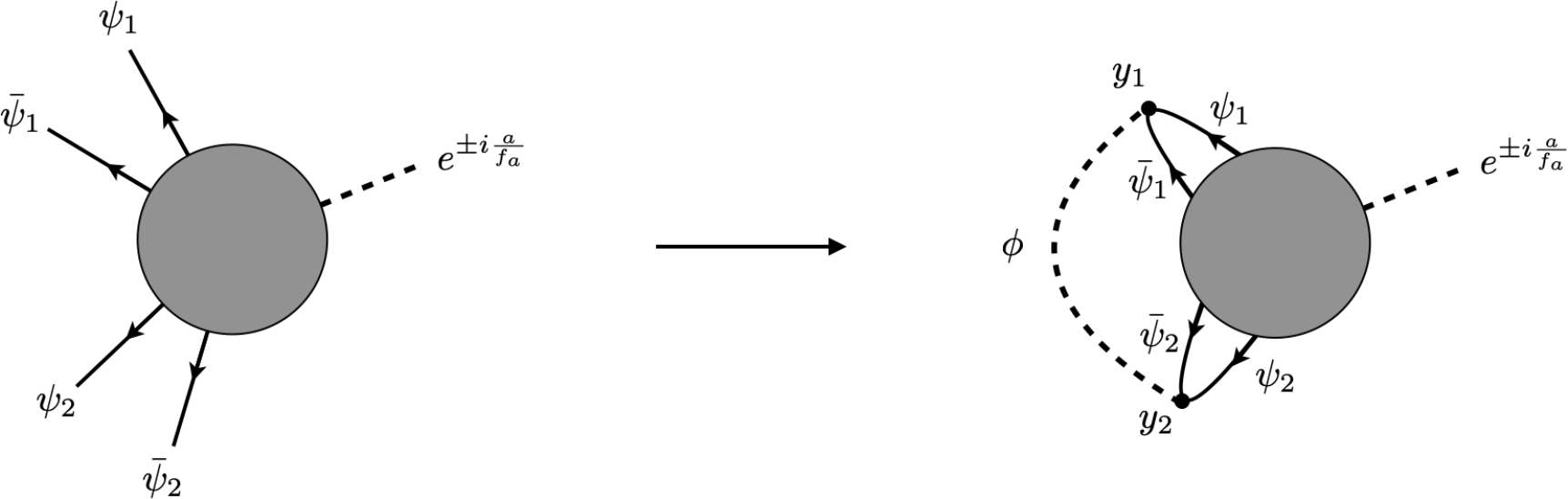}}} 
\ee
    \item Combine the couplings used to close the 't Hooft operator with the instanton density and the integral over the instanton size, which corresponds to an  overall factor 
    \be \label{eq:instMeasure}
    C_N \left(\frac{8\pi^2}{g^2}\right)^{2N} \int \frac{d\rho}{\rho^5} (\Lambda_G \rho)^{b_0}
    \ee
    and add appropriate powers of $\rho$, the only available dimensionful parameter, to get the right dimensions (i.e. those of a potential $V$). Propagators should not be added to the estimate. The integral over their momenta is cut off at $1/\rho$ in the instanton background (see Section~\ref{sec:full}) such that adding the appropriate powers of $\rho$ to get the right dimension automatically takes propagators of light particles into account. The instanton density for $SU(N)$, i.e. $C_N$, is given by ~\cite{tHooft:1976snw,Bernard:1979qt}
\begin{equation} \label{eq:CN}
C_N=\frac{K_1 e^{-(S^{(1/2)}- F^{(1/2)}) \alpha(1 / 2)-(S^{(1)}- F^{(1)}) \alpha(1)}}{(N-1) !(N-2) !}  e^{-K_2 N}\,,
\end{equation}
with $K_1\approx 0.466$, $K_2 \approx 1.678$ and $\alpha(0)=0,\alpha(1/2)=0.145873,\alpha (1) = 0.443307$. Numerical values for the function $\alpha (t)$ for different isospin representations $t$ can be found in~\cite{tHooft:1976snw}. $S^{(t)}$ and $F^{(t)}$ are the number of scalars and fermions which transform in the isospin $t$ representation under the $SU(2)$ containing the instanton. 
For instance the closed 't~Hooft vertex in Eq.~\eqref{eq:step2} is associated with an expression of the form
\be \label{eq:step3}
\vcenter{\hbox{\includegraphics[width=0.25\textwidth]{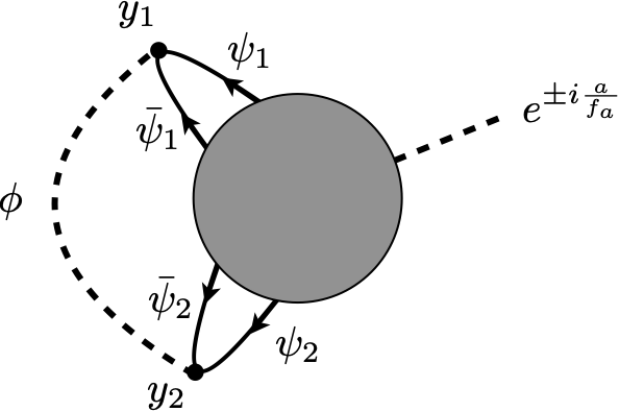}}} \sim C_N \left(\frac{8\pi^2}{g^2}\right)^{2N} e^{\pm i \frac{a}{f_a}} \int \frac{d\rho}{\rho^5} (\Lambda_{SU(N)} \rho)^{b_0} (y_1 y_2)
\ee
    \item Add a factor of $(4\pi)^{-\alpha}$, where $\alpha$ is given by
    \be \label{eq:LoopFactor}
    \alpha = \#\text{ fermion zero-modes} - 2\cdot \#\text{ vertices} + 2\cdot \#\text{ propagators}\, .
    \ee
    The vertices do not include the 't Hooft vertex and propagators do not include the fermion legs that exit it. Mass insertions count as a vertex, such that there is no loop factor associated with a 't Hooft operator closed exclusively with mass insertions. See Appendix~\ref{app:LoopFactor} for a derivation of this rule. The reason why the naive loop factor counting needs to be modified is that the fermion legs in the 't Hooft operator stand for zero-mode wavefunctions in the instanton background that contain explicit factors of $\pi$ and $\rho$.
        Including the $(4\pi)$ factors Eq.~\eqref{eq:step3} finally becomes
\be \label{eq:step4}
\vcenter{\hbox{\includegraphics[width=0.25\textwidth]{Figures/NDA_step3.pdf}}} \sim C_N \left(\frac{8\pi^2}{g^2}\right)^{2N} e^{\pm i \frac{a}{f_a}} \int \frac{d\rho}{\rho^5} (\Lambda_{SU(N)} \rho)^{b_0} \frac{y_1 y_2}{(4\pi)^2}
\ee
    \item Perform the integral over the instanton size $\rho$ and evaluate the gauge coupling in the prefactor at the scale where the integral over the instanton size is dominated. The running of the coupling in the pre-factor is a two-loop effect in the gauge coupling that one can neglect when performing our estimates. The limits of integration over the instanton size $\rho$ are $[M_{\rm UV}^{-1}, \infty]$.  $M_{\rm UV}$ is the UV cutoff of the EFT where we are performing the calculation. In the case of asymptotically-free theories we also need an IR cutoff $ M_{\rm IR}^{-1}$. We discuss how to choose $M_{\rm IR}$ in the rest of Section~\ref{sec:InstantonCalculus} and comment on $M_{\rm UV}$ at the end of this Section.
\end{enumerate}
Depending on the particle content and the couplings needed to close the 't Hooft operator, the integral can be UV or IR dominated. If the diagram contains loops the IR cutoff is the mass of the particle propagating in the loop. At lower energies the particle propagating in the loop should be integrated out. The 't Hooft vertex can still be closed with the resulting effective operator.

In asymptotically free gauge theories the integral over the instanton size is in general IR divergent. However, a fully reliable instanton calculation is still possible if the gauge group is completely broken by the VEV of a scalar field $\Phi$. In this scenario large instantons of size $g\rho \langle\Phi\rangle \gg 1$ are exponentially suppressed and the computation can be performed in the constrained instanton formalism~\cite{Affleck:1980mp}. The exponential suppression originates from the classical action of the scalar field in the instanton background $\delta S_0^\Phi \sim 2\pi^2\rho^2\langle\Phi\rangle^2$. Thus we can add a step $6$ to our recipe
\begin{enumerate}
    \item[$6$.] If the gauge group is completely broken by a scalar VEV of magnitude $v$, add a factor $e^{-2\pi^2 \rho^2 v^2}$ to the integrand and then integrate over the instanton size. This also regulates the IR divergence in asympotically  free gauge theories. If the group is partially broken one has to separately account for the instantons in the broken group which have the $e^{-2\pi^2 \rho^2 v^2}$ exponential suppression, and the instantons in the unbroken group without such a suppression. In such case additional $\mathcal{O}(1)$ factors will arise from accounting for the group theory factors of rotating the broken group inside the full group. 
\end{enumerate}

The above rules give a relatively precise estimate which differs from the full calculation at most by a factor of a few. Typically there is not a unique way to close the legs of the t'Hooft operator, and the above rules can help to identify the dominant contribution. In the next section we show how to apply these rules to toy examples that illustrate all cases of physical relevance.

Note that there is at least one case where our rules could be off by more than $\mathcal{O}(1)$ from the full result.\footnote{We thank Pablo Sesma for pointing this out to us.} This could happen if the scalar in the loop used to close up the 't Hooft operator's legs is itself charged under the gauge group of the instanton. In this case one would need to use the scalar propagator in the instanton background which is known analytically only in certain limiting cases~\cite{Brown:1977eb, Brown:1978bta}, and could lead to modifications of our rules for this case.

It is also useful to comment on the role of the UV cutoff $M_{\rm UV}$. In many theories the integral is exponentially convergent $\sim e^{-2\pi/\alpha(\rho)}$ and one can formally take $M_{\rm UV}\to\infty$ without affecting the computation. This is the case in the SM when computing QCD instantons. However, we find it more physical to adopt a Wilsonian point of view and always impose a UV cutoff.

Whenever Instanton NDA gives a UV-dominated result this should be taken with a grain of salt, as is always the case for UV-sensitive quantities in a EFT. It is possible, as in Section~\ref{sec:GUTs}, that at $M_{\rm UV}$ the axion potential receives instanton contributions from a larger gauge group and the estimate in the EFT is incorrect by large numerical factors. In these cases one has to apply Instanton NDA also in the UV theory to get a reliable estimate.

\subsection{Examples of Instanton NDA}

There are three physically distinct cases: we can close the legs of the t'Hooft operator with 1) relevant, 2) marginal or 3) irrelevant interactions of the fermions.
We now go through three examples that illustrate these three possibilities and make the application of our power-counting rules more concrete. 

We consider instanton contributions within an energy range $[M_{\rm IR}, M_{\rm UV}]$ which is well above the strong coupling scale $M_{\rm IR} \gg \Lambda_{G}$. This restriction ensures that instanton effects are calculable even when they are IR dominated. In the following we consider a $SU(N)$ gauge theory with two flavors of vector like fermions in the fundamental representation $\psi_1,\bar{\psi}_1$ and $\psi_2,\bar{\psi}_2$. In this case the 't Hooft operator has four legs, one for each fermion ($2k=1$ for the fundamental representation). The simplest way to close the 't Hooft operator is to use a relevant operator, in particular fermion mass insertions. This, according to the above rules, yields 
\be \label{eq:InstantonMass}
\begin{split}
\vcenter{\hbox{\includegraphics[width=0.23\textwidth]{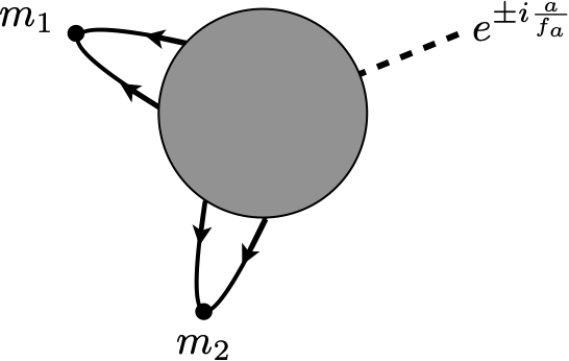}}} &\sim C_N \left( \frac{8\pi^2}{g^2} \right)^{2N} e^{\pm i \frac{a}{f_a}}\int_{1/M_{\rm UV}}^{1/M_{\rm IR}}\frac{d\rho}{\rho^5} (\Lambda_{SU(N)} \rho)^{b_0} m_1 m_2 \rho^2\\
&\sim C_N \left( \frac{8\pi^2}{g^2 (M_{\rm IR})} \right)^{2N} e^{\pm i \frac{a}{f_a}} \frac{m_1 m_2}{M_{\rm IR}^2}\left(\frac{\Lambda_{SU(N)}}{M_{\rm IR}}\right)^{b_0 -4} \Lambda_{SU(N)}^4\quad \text{for}\quad b_0 >2\,,
\end{split}
\ee
where we assumed that the integral is IR dominated, which is the case for $b_0>2$, and introduced IR and UV cutoffs $M_{\rm IR}$ and $M_{\rm UV}$, respectively.\footnote{For $SU(N)$ the beta function coefficient is given by $b_0 = \frac{11}{3}N	- \frac{2}{3} \sum_F T(F) d(F) - \frac{1}{3} \sum_S T(S) d(S)$, where the sum is over fermions $(F)$ and complex scalars $(S)$, $T$ is the Dynkin index and $d$ the dimension of the representation under further gauge groups.} Note that in order for the estimate to make sense $M_{\rm IR}\gg \Lambda_{SU(N)}$, such that the coupling is still perturbative. $\Lambda_{SU(N)}$ is the RG invariant scale defined in Eq.~\eqref{eq:RGI}.

The second general possibility is to use a marginal interaction. If the theory contains a scalar $\phi$ with Yukawa couplings to the fermions of the form $y_i \phi \psi_i \bar{\psi}_i$ one can also use Yukawa couplings and $\phi$ loops to close the 't Hooft vertex
\be \label{eq:InstantonHiggs}
\begin{split}
\vcenter{\hbox{\includegraphics[width=0.25\textwidth]{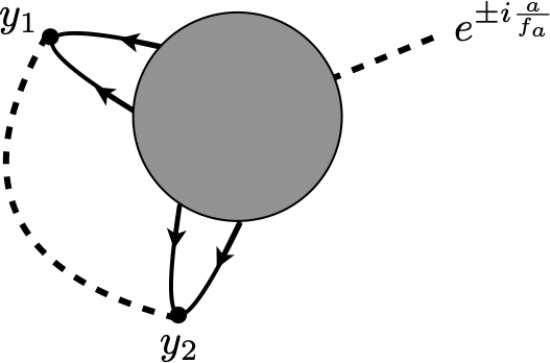}}} &\sim C_N \left( \frac{8\pi^2}{g^2} \right)^{2N} e^{\pm i \frac{a}{f_a}}\int_{1/M_{\rm UV}}^{1/M_{\rm IR}}\frac{d\rho}{\rho^5} (\Lambda_{SU(N)} \rho)^{b_0} \frac{y_1 y_2}{16\pi^2}\\
&\sim C_N \left( \frac{8\pi^2}{g^2(M_{\rm IR})} \right)^{2N} e^{\pm i \frac{a}{f_a}} \frac{y_1 y_2}{16\pi^2}\left(\frac{\Lambda_{SU(N)}}{M_{\rm IR}}\right)^{b_0 -4} \Lambda_{SU(N)}^4\quad \text{for}\quad b_0 >4\,,
\end{split}
\ee
where we again assumed that the integral is IR dominated, which is the case for $b_0 > 4$, and took $M_{\rm IR} > m_\phi$ such that $\phi$ is a propagating degree of freedom all the way to the IR cutoff. 
For instantons of size $1/\rho \in [M_{\rm IR}, M_{\rm UV}]$ this contribution dominates over the mass insertions in Eq.~\eqref{eq:InstantonMass} if $m_1 m_2 / M_{\rm IR}^2 < y_1 y_2/(16\pi^2)$. In theories where the scalar generates the fermion masses by obtaining a VEV $\langle \phi \rangle = v$ the contribution from $\phi$ loops dominates for $M_{\rm IR} > 4\pi v$.

The couplings used to close the legs of the 't Hooft operator do not have to be masses or marginal couplings but can also be higher-dimensional effective operators. Insertions of effective operators make the integral over the instanton size more UV dominated since powers of $\rho$ in the integrand are replaced by inverse powers of the EFT scale in the higher-dimensional operator, giving a smaller power of $\rho$. In the current example we  use a four-fermion operator of the form
\be \label{eq:fourFermion}
\frac{c_F}{\Lambda_F^2}\psi_1 \bar{\psi}_1 \psi_2 \bar{\psi}_2
\ee
to close all legs of the 't Hooft vertex
\be \label{eq:InstantonEffOperator}
\begin{split}
\vcenter{\hbox{\includegraphics[width=0.25\textwidth]{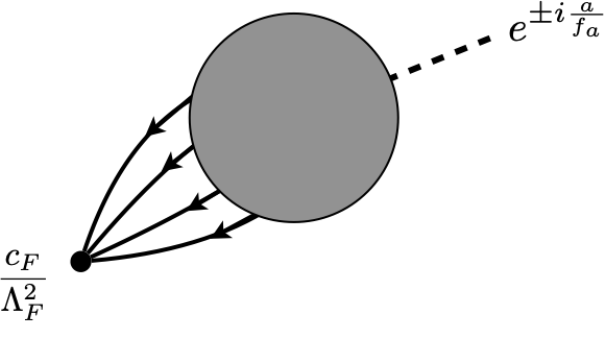}}} &\sim C_N \left( \frac{8\pi^2}{g^2} \right)^{2N} e^{\pm i \frac{a}{f_a}}\int_{1/M_{\rm UV}}^{1/M_{\rm IR}}\frac{d\rho}{\rho^5} (\Lambda_{SU(N)} \rho)^{b_0} \frac{c_F}{(4\pi)^2\rho^2\Lambda_F^2}\,,
\end{split}
\ee
%
which is UV dominated for $b_0 < 6$. Note that we have also included a factor of $1/(4\pi)^2$ according to rule number 4 in the previous section. This is also required for UV dominated instanton effects within the EFT to match IR dominated instanton effects in the UV theory. As an example consider the above theory with two vector like fermions in the fundamental representation and a real scalar $\phi$ with Yukawa couplings $y_i \phi \psi_i\bar{\psi}_i$ and assume that $4 < b_0 < 6$. At energies where $\phi$ is a propagating degree of freedom we can close the 't Hooft operator with $\phi$ loops which yields Eq.~\eqref{eq:InstantonHiggs}. At energies $E\ll m_\phi$ we can integrate out $\phi$ and obtain an effective operator of the form $\frac{y_1 y_2}{m_\phi^2} \psi_1\bar{\psi}_1 \psi_2\bar{\psi}_2$. If we use this operator to close the 't Hooft operator we find Eq.~\eqref{eq:InstantonEffOperator} which is UV dominated and gives
\be
\begin{split}
\vcenter{\hbox{\includegraphics[width=0.25\textwidth]{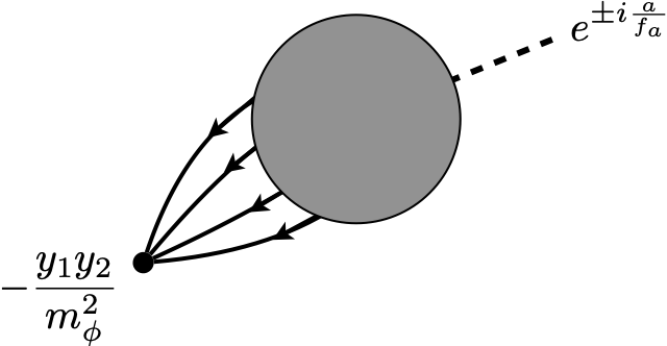}}} &\sim C_N \left( \frac{8\pi^2}{g^2} \right)^{2N} e^{\pm i \frac{a}{f_a}}\int_{1/M_{\rm UV}}^{1/M_{\rm IR}}\frac{d\rho}{\rho^5} (\Lambda_{SU(N)} \rho)^{b_0} \left(\frac{y_1 y_2}{(4\pi)^2\rho^2 m_\phi^2}\right)\\
&\sim C_N \left( \frac{8\pi^2}{g^2 (M_{\rm UV})} \right)^{2N} e^{\pm i \frac{a}{f_a}} \frac{y_1 y_2}{16\pi^2} \frac{M_{\rm UV}^2}{m_\phi^2} \left(\frac{\Lambda_{SU(N)}}{M_{\rm UV}}\right)^{b_0 - 4} \Lambda_{SU(N)}^4\,,
\end{split}
\ee
which for the natural choice of $M_{\rm UV} = M_{\rm IR} = m_\phi$ gives parametrically identical contributions to the potential in the EFT and UV theory.
%
\subsection{Axion Potential from Low-energy QCD} \label{sec:lowE}
%
In the previous Section we outlined how to estimate instanton contributions to the axion potential. However, in order to make meaningful statements we always had to introduce an IR cutoff for the instanton size integration. This was necessary to stay within the perturbative regime, where instantons are the dominant saddle of the path integral.

If the contribution to the potential is dominated by large IR instantons, which is typically the case in asymptotically free gauge theories, it is tempting to extrapolate the estimate into the non-perturbative region by taking $M_{\rm IR}\rightarrow \Lambda_{\rm QCD}$. However, in this region multi-instanton effects are not suppressed, implying that any attempt of finding contributions to the axion potential from closing legs of 't Hooft operators at the strong coupling scale is not very meaningful. In fact arguments from large $N$ QCD~\cite{Witten:1978bc,Veneziano:1979ec,Witten:1980sp} and supersymmetric QCD~\cite{Dine:2016sgq,Csaki:2023yas} suggest confinement dynamics and not large instantons give the dominant contribution to the axion and $\eta'$ potentials. Let us however stress that this does not rule out the existence of contributions to the potential from small UV instantons which can be estimated according to the rules in the previous section.

The appropriate way to determine the axion potential from low-energy QCD is within chiral perturbation theory that yields~\cite{GrillidiCortona:2015jxo}
\be
V_{\rm IR}(a)=-m_\pi^2 f_\pi^2 \sqrt{1-\frac{4 m_u m_d}{\left(m_u+m_d\right)^2} \sin ^2\left(\frac{a}{2 f_a}\right)}\,.
\label{eq:VIR}
\ee
One can then use Instanton NDA to compare this irreducible IR contribution with the model-dependent UV-instantons of choice. In the perturbative regime UV contributions are always exponentially suppressed by the one-instanton action $V_{\rm UV}(a)/V_{\rm IR}(a)\sim e^{-2\pi/\alpha(M_{\rm UV})}$.
Therefore, it is legitimate to wonder if the UV contribution can ever dominate over the IR one when the UV instanton calculation is under pertubative control. In practice you might wonder if our rules are ever useful at all. 
However, performing the  Instanton NDA estimate, including dimensionful factors, one can see immediately that the ratio $V_{\rm UV}(a)/V_{\rm IR}(a)\sim (M_{\rm UV}/\Lambda_{\rm QCD})^4 e^{-2\pi/\alpha(M_{\rm UV})}$ can easily be larger than one, even if $\alpha(M_{\rm UV})$ is perturbative. This is discussed in more detail  in the first paragraph of Section~\ref{sec:EnhancedSmallInstantons}.

In the next Section we make contact between Instanton NDA and full instanton calculations. We then move on to examples of physical relevance.
%
\subsection{Full Instanton Computation}
\label{sec:full}
%
Let us now outline how to turn the estimates and power counting rules of the previous sections into a fully-fledged instanton calculation which includes all $\mathcal{O}(1)$ factors, following the conventions of~\cite{Csaki:2019vte}. For more details and background information see~\cite{Shifman:2012zz,Csaki:2019vte,Ruhdorfer:2021xcb}. We would also like to remind the reader that all calculations are performed in Euclidean spacetime.

In order to compute contributions to the axion potential one has to find the vacuum-to-vacuum transition amplitude in the one-instanton and axion background. As we have outlined in Section~\ref{sec:InstPowerCounting} this is done by expanding the fields around the classical instanton background and performing the quadratic part of the path integral over the quantum fluctuations with the integral over zero-modes separated out. For an $SU(N)$ gauge theory the resulting vaccum-to-vacuum amplitude takes the form
\begin{equation}\label{eq:SUNInstanton}
	W_{S U(N)}= C_N \left(\frac{8 \pi^2}{g^2}\right)^{2 N}\int d \kappa \int \frac{d^4 x_0 d \rho}{\rho^5} e^{-\frac{8 \pi^2}{g^2(1 / \rho)}-i\frac{a}{f_a}} e^{-S_{\rm scalar}(\kappa)} \int \prod_{f} \rho^{1 / 2} d \xi_f^{(0)}\,,
\end{equation}
where the integral is over all fermion zero modes $\xi_f^{(0)}$ and the instanton collective coordinates: the instanton location $x_0$, the instanton size $\rho$ and the orientation within the gauge group, denoted by the normalized integral over $\kappa$. Note that, as already mentioned, the gauge coupling in the prefactor comes without any explicit scale dependence as the running of the gauge coupling in the instanton measure is a two-loop effect. In the following we will always use the gauge coupling at the scale where the instanton size integration is dominated. We have also included a possible contribution to the classical action from a scalar charged under the gauge group which obtains a VEV, $S_{\rm scalar}(\kappa)$. In general this contribution depends on the orientation of the instanton in the gauge group w.r.t. the scalar VEV, which has to be taken into account when performing the $\int d\kappa$ integral. The numerical prefactor $C_N$ from the Jacobian associated to the transformation of zero modes to collective coordinates and the integral over the non-zero modes of all particles charged under the gauge group is given in Eq.~\eqref{eq:CN}.

Let us now elaborate further on how Eq.~\eqref{eq:SUNInstanton} is connected to closing legs of the 't Hooft operator. The integral $\int d\xi_f^{(0)}$ projects onto the zero mode wavefunction of fermion fields. For $\psi_f (x) = \sum_i \psi_f^{(i)} (x) \xi^{(i)}_f$  the integral yields $\int d\xi_f^{(0)} \psi_f(x) = \psi_f^{(0)}$, where $\psi_f^{(0)}$ is the zero mode wavefunction. For a Weyl fermion in the fundamental representation in singular gauge, where the topological charge of the instanton is saturated at the instanton center, this is given by~\cite{Shifman:2012zz}
\be\label{eq:FermZeroMode}
\psi_f^{(0)}(x)_{\alpha i}=\frac{\rho}{\pi} \frac{i \left(x-x_0\right)_\mu U^j_i \left(\tau_\mu^{+}\right)_j^k}{(\left(x-x_0\right)^2)^{1 / 2}(\left(x-x_0\right)^2+\rho^2)^{3 / 2}}   \varphi_{\alpha k}\,,
\ee
where $\varphi_{\alpha k} \sim \epsilon_{\alpha k}$ is a Weyl spinor, $\alpha, i,j,k$ are the spinor and $SU(N)$ indices, respectively, and $\tau_\mu^{\pm} = (\vec{\tau}, \mp i)$ with $\vec{\tau}$ being the vector of Pauli matrices. Note that $U \equiv U(\kappa) \in SU(N)$ parameterizes the arbitrary orientation of the instanton within the gauge group.\footnote{Note that $U$ typically drops out in the computation of the vacuum-to-vacuum amplitude since the couplings used to close the legs of the 't Hooft operator originate from gauge invariant operators.}

In order to soak up the zero modes one needs further interactions involving the fermions charged under the gauge group. The simplest example is to treat the mass term as an interaction. Going back to the example we considered in the previous sections, i.e. an $SU(N)$ gauge group with two vector like fermions in the fundamental representation, we can add a mass term for both fermions in the action
\be
- S_\psi = i \int d^4x\left( m_1 \bar{\psi}_1 \psi_1 + m_2 \bar{\psi}_2 \psi_2 \right)\,.
\ee
Adding this to Eq.~\eqref{eq:SUNInstanton} and performing the integral over the zero modes one finds
\be
\begin{split}
\int \rho^2 d\bar{\xi}^{(0)}_1 d\xi^{(0)}_1 d\bar{\xi}^{(0)}_2 d\xi_2^{(0)} e^{-S_\psi} &= -\rho^2 \int d^4 x_1 m_1 \bar{\psi}^{(0)}_1 (x_1) \psi^{(0)}_1 (x_1) \int d^4 x_2 m_2 \bar{\psi}^{(0)}_2 (x_2) \psi^{(0)}_2 (x_2)\\
&=-\rho^2 m_1 m_2\,,
\end{split}
\ee
where we used that the zero-mode wavefunction is normalized. This exactly reproduces the power counting rules and confirms that closing legs of the 't Hooft operator with mass insertions is not a loop.

When using propagating particles in loops to close the operator one has to include them and their interaction in $W_{SU(N)}$. Let us assume that the theory contains a neutral scalar $\phi$ with Yukawa couplings to the fermions. In this case $W_{SU(N)}$ is of the form
\begin{equation}\label{eq:InstantonYukawa}
\begin{split}
	W_{S U(N)}= C_N \left(\frac{8 \pi^2}{g^2}\right)^{2 N} \int d \kappa &\int \frac{d^4 x_0 d \rho}{\rho^5} e^{-\frac{8 \pi^2}{g^2(1 / \rho)}-i\frac{a}{f_a}} e^{-S_{\rm scalar}(\kappa)} \int \mathcal{D}\phi\, e^{-S_0[\phi]} \\
 &\times \int \prod_{i=1}^2 \rho\, d \bar{\xi}_i^{(0)} d\xi_i^{(0)} e^{i\int d^4 x \sum_i y_i \phi(x) \bar{\psi}_i(x) \psi_i (x)}\,,
 \end{split}
\end{equation}
where $S_0[\phi]$ is the free action for the scalar $\phi$. Performing the integral over the fermion zero modes and the functional integral over $\phi$ one finds
\bea
\mathcal{I} &=& \int \mathcal{D}\phi\, e^{-S_0[\phi]} \prod_{i=1}^2 \left( i y_i \rho \int d^4 x_i\, \phi (x_i) \bar{\psi}_i^{(0)} (x_i) \psi_i^{(0)} (x_i) \right) \nn \\
&=& -y_1 y_2 \rho^2 \int d^4 x_1 \int d^4 x_2 \bar{\psi}_1^{(0)} (x_1) \psi_1^{(0)} (x_1) \bar{\psi}_2^{(0)} (x_2) \psi_2^{(0)} (x_2) \Delta_F (x_1 -x_2)\,,
\eea
where the integral over the fermion zero modes picks the second order term in the expansion of the exponential containing the Yukawa interaction. The path integral over $\phi$ gives a Feynman propagator $\Delta_F (x_1 -x_2)$. Using the explicit expression of the zero modes in Eq.~\eqref{eq:FermZeroMode} and the usual scalar propagator this evaluates to
\be
\mathcal{I} = y_1 y_2\rho^2 \int \frac{d^4p}{(2\pi)^4} \frac{(p\rho)^2 K_1^2(p\rho)}{p^2 + m_\phi^2} =\begin{cases}
    \frac{y_1 y_2}{5\pi^2}\frac{1}{\rho^2 m_\phi^2}\,,\qquad \rho m_\phi \gg 1\\
    \frac{y_1 y_2}{12\pi^2}\,,\qquad\qquad \rho m_\phi \ll 1
\end{cases}\,,
\ee
where $K_1$ is a modified Bessel function of the second kind. Thus the loop momentum integral is cut off at $p\sim 1/\rho$. For $1/\rho \ll m_\phi$ this essentially corresponds to integrating out the scalar. The result is consistent with the power counting rules which would predict $\mathcal{I}=\tfrac{y_1 y_2}{16\pi^2}$ for $1/\rho \gg m_\phi$.

In order to obtain the axion potential from the vacuum-to-vacuum amplitude $W_{SU(N)}$ one has to sum over all instanton and anti-instanton configurations. In the dilute instanton gas approximation~\cite{Callan:1977gz} one assumes that the dominant contribution to this sum originates from well-separated, non-interacting instantons and anti-instantons. This gives an effective axion potential of the form
\be
e^{-\int d^4 x\, V(a)}
\approx \sum_{n_{+}, n_{-}=0}^{\infty} \frac{1}{n_{+} !} \frac{1}{n_{-} !} W_{S U(N)}^{n_{+}} \bar{W}_{S U(N)}^{n_{-}}  = e^{ W_{S U(N)}+\bar{W}_{S U(N)}}\,,
\ee
where $n_+$ and $n_-$ are the number of instantons and anti-instantons and $W_{S U(N)}$ and $\bar{W}_{S U(N)}$ are the vacuum-to-vacuum amplitudes in the instanton and anti-instanton background, respectively. The dilute instanton gas description is a good approximation for $\exp [ -\tfrac{8\pi^2}{g^2} ] \ll 1$  and breaks down when the theory becomes strongly coupled.

%
\section{Applications of Instanton NDA}\label{sec:examples}
%
In this section we apply the instanton power counting rules introduced in Section~\ref{sec:InstantonCalculus} to illustrate a few mechanisms to modify the axion mass using UV instantons which have been proposed in the literature~\cite{Holdom:1982ex,Holdom:1985vx,Flynn:1987rs,Gherghetta:2020keg,Agrawal:2017ksf,Csaki:2019vte,Dine:1986bg,Choi:1998ep,Rubakov:1997vp}. We additionally comment on misaligned contributions to the axion potential which may enter the instanton calculation through CP violating higher-dimensional operators.
%
\subsection{Enhanced UV Instanton Contributions to the Axion Potential} \label{sec:EnhancedSmallInstantons}
%
The absolute size of instanton contributions is determined by a combination of the running coupling and the energy scale one is sensitive to in the integral over the instanton size, i.e. if one uses only marginal couplings to close the legs of the 't Hooft operator the size of the instanton contribution is roughly given by $\max_M \exp(-\tfrac{8\pi^2}{g^2(M)})M^4$. This implies that UV instantons get more important if the running to smaller couplings in the UV is slower than the increase in the scale, which naively occurs for $b_0 < 4$. If dimensionful couplings close the legs of the 't Hooft operator there are additional  suppression factors which are powers of $m_\psi/M$ for insertions of light fermion masses ($m_\psi < M$) or $M/M_{\rm UV}$ where $M_{\rm UV} \geq M$ is the suppression scale of higher-dimensional operators. This modifies the above estimate for the size of the instanton contribution to
\be
m_{\psi}^{n_\psi} M_{\rm UV}^{-n_{\rm UV}} \cdot \max_M\, M^{4-n_\psi + n_{\rm UV}} e^{-\frac{8\pi^2}{g^2(M)}}\,,
\ee
where $n_\psi$ is the number of mass insertions and $n_{\rm UV}$ the power of suppression scales from higher-dimensional operators. This implies that the instanton contribution is UV dominated if $b_0 < 4-n_\psi + n_{\rm UV}$. However, it is important to keep in mind that even if a particular instanton configuration is UV dominated this does not necessarily imply that there are no other IR contributions which are larger. 

Any attempt to make UV instantons more important requires an effective reduction of the beta function in the UV. There are various ways to achieve this which have been proposed in the literature. The simplest possibility is to add additional matter charged under the gauge group (see e.g.~\cite{Holdom:1982ex,Holdom:1985vx,Flynn:1987rs}). Note however, that if the additional particles are fermions, these have zero modes in the instanton background such that there are additional legs in the 't~Hooft vertex that have to be closed. If those legs are closed with mass insertions this will suppress UV instantons (see below for an example). This suppression may be overcome if the new fermions have Yukawa couplings that can be used to close the legs with scalar loops. 

Let us now see how this enhancement of small instantons can be seen with the power counting rules of the previous section. We again work with an $SU(N)$ gauge theory with two vector like fermions in the fundamental representation and assume for simplicity that all additional particles responsible for reducing the beta function are scalars with a mass $M$. Then the RGE invariant scale of the low energy theory $\Lambda_{SU(N)}^{\rm IR} < M$ is related to the one of the UV theory $\Lambda_{SU(N)}$ (which contains the additional particles), through the matching relation
\be \label{eq:ScaleMatching}
\left(\frac{\Lambda_{SU(N)}^{\rm IR}}{M}\right)^{b_0^{\rm IR}} = \left(\frac{\Lambda_{SU(N)}}{M}\right)^{b_0}\,.
\ee
Note that $\Lambda_{SU(N)}^{\rm IR}$ is fixed given  knowledge of the low-energy theory. In the following we will also take $(\Lambda_{SU(N)}^{\rm IR})^4$ as the typical size of IR strong dynamics effects and compare that to the size of UV instanton contributions.

We can now go through the different possibilities to close the 't Hooft operators that we outlined in the previous section and use the matching relation in Eq.~\eqref{eq:ScaleMatching} to compare it to the IR contribution. Whenever the integral over the instanton size is IR dominated we take $M_{\rm IR} = M$. Closing the 't Hooft operator with mass insertions we find
\be \label{eq:InstantonMassUV}
\begin{split}
\vcenter{\hbox{\includegraphics[width=0.25\textwidth]{Figures/massPotential.pdf}}}\\ \sim C_N  \left( \frac{8\pi^2}{g^2 (M_{\rm UV})} \right)^{2N} &e^{\pm i \frac{a}{f_a}} (\Lambda_{SU(N)}^{\rm IR})^4
\times
\begin{cases}
   \left(\tfrac{g^2 (M_{\rm UV})}{g^2(M)}\right)^{2N}\frac{m_1 m_2}{M^2}  \left(\frac{\Lambda_{SU(N)}^{\rm IR}}{M}\right)^{b_0^{\rm IR} -4}\,,\,b_0 >2\\
    \frac{m_1 m_2}{M^2} \left(\frac{M_{\rm UV}}{M}\right)^{2-b_0} \left(\frac{\Lambda_{SU(N)}^{\rm IR}}{M}\right)^{b_0^{\rm IR}-4}\,,\quad b_0 < 2
\end{cases}
\end{split}
\ee
Mass insertions make the instanton size integral more IR dominated because the masses enter the instanton calculation in the combination $(\rho m_\psi)$ which grows in the IR. Thus a stronger reduction of the beta function is needed to make instantons UV-dominated. For $b_0 > 2$ the instanton is IR dominated and all {numerical} factors {that multiply the dimensionful quantity $(\Lambda_{SU(N)}^{\rm IR})^4$ in Eq.~\eqref{eq:InstantonMassUV}} are strictly smaller than one, such that this contribution is always smaller than the contribution from IR strong dynamics effects which scales like $(\Lambda_{SU(N)}^{\rm IR})^4$. In contrast for $b_0 < 2$ the instanton is UV dominated since $(M_{\rm UV}/M)^{2-b_0}$ becomes an enhancemant instead of a suppression. This means that UV instanton contributions can be larger than contributions from IR strong dynamics if $M_{\rm UV} \gg M$ and $M \sim \Lambda_{SU(N)}^{\rm IR}$, but there is an irreducible suppression of $m_1 m_2 / M^2$ and the beta function has to be extremely small, $b_0 < 2$. For reference, in QCD $b_0=7$ above the top mass, and at least 4 new quark flavors are required for $b_0\leq2$.

Closing the 't Hooft operator with Yukawa couplings and loops of scalars resolves both problems
\be \label{eq:InstantonHiggsUV}
\begin{split}
\vcenter{\hbox{\includegraphics[width=0.25\textwidth]{Figures/HiggsPotential.pdf}}}\\ \sim C_N \left( \frac{8\pi^2}{g^2 (M_{\rm UV})} \right)^{2N} &e^{\pm i \frac{a}{f_a}} (\Lambda_{SU(N)}^{\rm IR})^4 \times 
\begin{cases}
    \frac{y_1 y_2}{16\pi^2}\left(\tfrac{g^2(M_{\rm UV})}{g^2(M)}\right)^{2N}\left(\frac{\Lambda_{SU(N)}^{\rm IR}}{M}\right)^{b_0^{\rm IR} -4}\,, \, b_0 >4\\
     \frac{y_1 y_2}{16\pi^2}\left( \frac{M_{\rm UV}}{M} \right)^{4-b_0}\left(\frac{\Lambda_{SU(N)}^{\rm IR}}{M}\right)^{b_0^{\rm IR} -4}\,,  \,\,\,\,\, b_0 <4
\end{cases}
\end{split}
\ee
there is no mass suppression and one obtains an enhancement by powers of $M_{\rm UV}/M \gg 1$ already for $b_0 < 4$.

Using the four fermion operator in Eq.~\eqref{eq:fourFermion} to close the legs yields a parametrically similar result with the main difference that the integral over instanton sizes is UV dominated already for $b_0 < 6$
\be \label{eq:InstEffOp}
\begin{split}
\vcenter{\hbox{\includegraphics[width=0.25\textwidth]{Figures/effOperatorPotential.pdf}}} \sim C_N &\left( \frac{8\pi^2}{g^2 (M_{\rm UV})} \right)^{2N} e^{\pm i \frac{a}{f_a}} (\Lambda_{SU(N)}^{\rm IR})^4 \\
&\times
\begin{cases}
\frac{c_F}{(4\pi)^2} \left( \tfrac{g^2(M_{\rm UV})}{g^2(M)} \right)^{2N} \left(\frac{M}{\Lambda_F}\right)^2 \left(\frac{\Lambda_{SU(N)}^{\rm IR}}{M}\right)^{b_0^{\rm IR} -4} \,, \qquad b_0 >6\\
 \frac{c_F}{(4\pi)^2}   \left( \frac{M}{\Lambda_F} \right)^{2} \left( \frac{M_{\rm UV}}{M} \right)^{6-b_0}
\left(\frac{\Lambda_{SU(N)}^{\rm IR}}{M}\right)^{b_0^{\rm IR} -4}\,, \qquad\, 4 < b_0 <6\\
\frac{c_F}{(4\pi)^2}   \left( \frac{M_{\rm UV}}{\Lambda_F} \right)^{2} \left( \frac{M_{\rm UV}}{M} \right)^{4-b_0}
\left(\frac{\Lambda_{SU(N)}^{\rm IR}}{M}\right)^{b_0^{\rm IR} -4}\,, \qquad\,  b_0 <4
\end{cases}
\end{split}
\ee
Once the instanton contribution is UV dominated the $(M/\Lambda_F)^2$ suppression gets gradually turned into $(M_{\rm UV}/\Lambda_F)^2$ for $4< b_0 <6$. For $b_0 < 4$ the suppression essentially vanishes since one expects $\Lambda_F \sim M_{\rm UV}$ and one obtains an additional enhancement from the $(M_{\rm UV}/M)^{4-b_0}$ factor. This implies that for $b_0 < 4$ contributions including insertions of higher-dimensional operators are not suppressed w.r.t. contributions which use only marginal couplings to close the legs of the 't Hoof operator. They can even be larger depending on the size of the Wilson coefficient $c_F$ relative to the Yukawa couplings.
This shows the surprising result that even instanton effects with effective operators can lead to contributions to the axion potential which are larger than the one from low-energy non-perturbative dynamics. 

These examples also illustrate why $SU(2)_L$ instantons are UV dominated when the SM is embedded in a Grand Unified Theory, while the axion potential from QCD remains IR dominated. $SU(2)_L$ instantons generate the 't Hooft operator $(qqq\ell)^3$ and can give an axion potential together with three insertions of the operator $(qqq\ell)/\Lambda_F^2$. The latter is generated by integrating out the triplet Higgs in GUTs. Three insertions make the instanton calculation strongly UV dominated, with an instanton measure scaling as $\sim \int (d\rho/\rho^5) (1/(\rho \Lambda_F)^6)$. The QCD instantons generate the operator $\prod_{i=1}^F \bar  q_i q_i$, whose legs can be closed with relevant interactions (quark masses) or marginal interactions (Yukawa couplings), giving rise to a IR dominated result (for QCD $b_0\leq 7$ at all scales).

The above discussion suggests that obtaining UV-dominated QCD axion potentials in the SM requires the introduction of a large set of new particles to modify the beta function. However there is another possibility: one can consider embeddings of QCD into a larger gauge group $G'$ in the UV with a non-trivial index of embedding (see e.g.~\cite{Intriligator:1994jr,Intriligator:1994sm,Intriligator:1995id,Csaki:1998vv,Csaki:2019vte}). In Appendix~\ref{app:InstIOE} we demonstrate how this mechanism can be understood with Instanton NDA.

Further possibilities to modify the axion potential from UV contributions include embedding the theory into an extra dimension~\cite{Gherghetta:2020keg} or coupling the axion to a new confining gauge group~\cite{Rubakov:1997vp,Fukuda:2015ana,Berezhiani:2000gh,Hook:2014cda,Blinov:2016kte,Dimopoulos:2016lvn,Gherghetta:2016fhp}, but we do not discuss these examples here.
%
\subsection{Misalignment from CP-violating Operators}\label{sec:misaligned}
%
In order to solve the strong CP problem the axion potential must have a minimum close to zero since a non-vanishing axion VEV induces an effective $\bar{\theta}$ angle, $\bar{\theta}_{\rm ind} = \langle a\rangle /f_a$, which is bounded to be $\bar{\theta}_{\rm ind} < 10^{-10}$ from neutron electric dipole moment measurements. It is well-known that higher-dimensional operators that break the PQ symmetry generate a contribution to the potential which misaligns the minimum from zero. Consider e.g. an effective operator of the form $\tfrac{c}{2 M^{n-4}}\Phi^n + {\rm h.c.}$ where $\Phi = f_a e^{i a/f_a}$ is the fundamental PQ scalar whose angular component can be identified with the axion. In combination with the potential generated from instantons or the strong dynamics, this takes the form
\be
V(a) = -\Lambda_{\rm QCD}^4 \cos\left( \frac{a}{f_a} + \bar{\theta}\right) + c f_a^4 \left(\frac{f_a}{M}\right)^{n-4} \cos\left(\frac{n\, a}{f_a}\right)\,.
\ee
To leading order in the effective operator the minimum of the potential is at
\be
\left\langle \frac{a}{f_a}\right\rangle = -\bar{\theta} -c\, n \left(\frac{M}{\Lambda_{\rm QCD}}\right)^4 \left(\frac{f_a}{M}\right)^n \sin (n\bar{\theta})\,.
\ee
If we take $c,\bar{\theta}\sim \mathcal{O}(1)$, even for Planck suppressed operators, i.e. $M=\Mpl$, and $f_a = 10^{10}$~GeV, one needs $n\gtrsim 10$ (see also~\cite{Holman:1992us}). This is the so-called PQ quality problem, i.e. the axion solution to the strong CP problem requires a high-quality global PQ symmetry. Quantum gravity is not expected to respect global symmetries such that Planck suppressed PQ-violating operators are generically present, but note that they could also exist with only a small pre-factor $c$, i.e. $c\sim e^{-S_{\rm cl}}$, and be an instanton-like non-perturbative effect.

However, even if there is a high-quality PQ symmetry and PQ breaking operators are absent up to a high operator dimension, a misaligned contribution to the axion potential can also be generated from instantons (see e.g.~\cite{Dine:2022mjw}). It is easy to see how this works. In Sections~\ref{sec:InstantonCalculus} and~\ref{sec:examples} we implicitly assumed that all couplings are real. However, if the couplings have a non-vanishing phase and violate CP, the phase will enter the axion potential.
In the example of the QCD axion this already happens within the SM where the CP violating phase of the Yukawa matrices produces a misalignment in the axion potential which is however far below the experimental sensitivity~\cite{Georgi:1986kr}. This becomes more relevant in the presence of higher-dimensional CP-violating operators which give a potentially measurable contribution to the neutron dipole moment~\cite{Bigi:1990kz,Pospelov:2000bw,Pospelov:2005pr}. To see how this works let us go back to our example in Section~\ref{sec:InstantonCalculus} with an $SU(N)$ gauge group and two vector like fermions in the fundamental representation. If we now close the fermion legs of the 't Hooft operator with an effective operator $\frac{c_F}{\Lambda_F^2}\psi_1 \bar{\psi}_1 \psi_2 \bar{\psi}_2 + {\rm h.c.}$ which has a complex coefficient, i.e. $c_F = |c_F| e^{i \delta_F}$ the contribution to the axion potential 
is of the form 
\be
\begin{split}
\vcenter{\hbox{\includegraphics[width=0.23\textwidth]{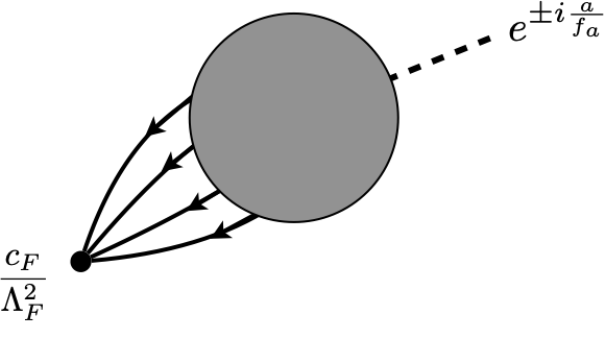}}} + \, {\rm h.c.}\,\sim\, \frac{|c_F|}{(4\pi)^2} C_N &\left(\frac{8\pi^2}{g^2(M_{\rm UV})}\right)^{2N} \Lambda_{SU(N)}^4 \cos\left(\frac{a}{f_a} + \delta_F\right)\\
&\times\begin{cases}
 \left( \tfrac{g^2(M_{\rm UV})}{g^2(M_{\rm IR})} \right)^{2N}\left(\frac{M_{\rm IR}}{\Lambda_F^2}\right)^2 \left(\frac{\Lambda_{SU(N)}}{M_{\rm IR}}\right)^{b_0 -4} \,, \quad\, b_0 >6\\
\left(\frac{M_{\rm UV}}{\Lambda_F^2}\right)^2\left(\frac{\Lambda_{SU(N)}}{M_{\rm UV}}\right)^{b_0 -4}\,, \,\,\, \qquad\qquad\qquad\,\, b_0 <6
\end{cases}
\end{split}
\ee
where we expect $\Lambda_F \sim M_{\rm UV}$.
If there is a mechanism that increases the axion mass through enhanced small instantons, these small instantons also enhance the effect of CP violating higher-dimensional operators as was pointed out in~\cite{Bedi:2022qrd}. Thus a successful axion solution of the strong CP problem does not only require a high-quality PQ symmetry but also the absence of CP violation up to a high scale. The vulnerability to these misaligned instanton contributions in the axion potential is however strongly model dependent. We discuss this in more detail in {the next} Section. 
%
\section{Misalignment from Instantons: A Large UV Axion Mass Increases  $\bar\theta$}\label{sec:misalignment}
%
As we have already mentioned in Section~\ref{sec:misaligned}, a high-quality PQ symmetry does not necessarily guarantee a successful axion solution to the strong CP problem. Instanton effects involving CP-violating higher-dimensional operators can generate a misaligned contribution to the axion potential which induces a non-vanishing effective QCD $\bar{\theta}$ angle.

In this section we use the insights and power counting rules developed in Section~\ref{sec:InstantonCalculus} to estimate such effects. Misaligned contributions to the axion potential can generically be either IR or UV dominated. IR dominated contributions can easily be estimated within a low-energy EFT whereas UV dominated contributions and their sizes are extremely model dependent and crucially rely on the particle spectrum and the running of the QCD coupling in the UV. As was already pointed out in~\cite{Bedi:2022qrd} mechanisms which enhance small instantons also boost the effect of CP violating operators in misaligning the axion potential. See also~\cite{Dine:2022mjw} for a related discussion mainly in the context of supersymmetric theories and Grand Unified Theories.
%
\subsection{Enhanced Small Instantons}\label{sec:SmallInstanton}
%
In scenarios where the axion mass is enhanced by small instantons the misaligned contribution to the axion potential from CP violating effective operators also gets boosted. 

Before putting the discussion into a broader context we go through two simple examples where CP violating effects mediated by UV dominated instantons play an important role: models with a reduced QCD beta function in the UV and possible misaligned contributions from $SU(2)_L$ instantons.
%
\subsubsection{Modified QCD Beta Function in the UV}\label{sec:QCDBetaFunction}
\begin{figure}[t]
	\centering
	\includegraphics[width=\textwidth]{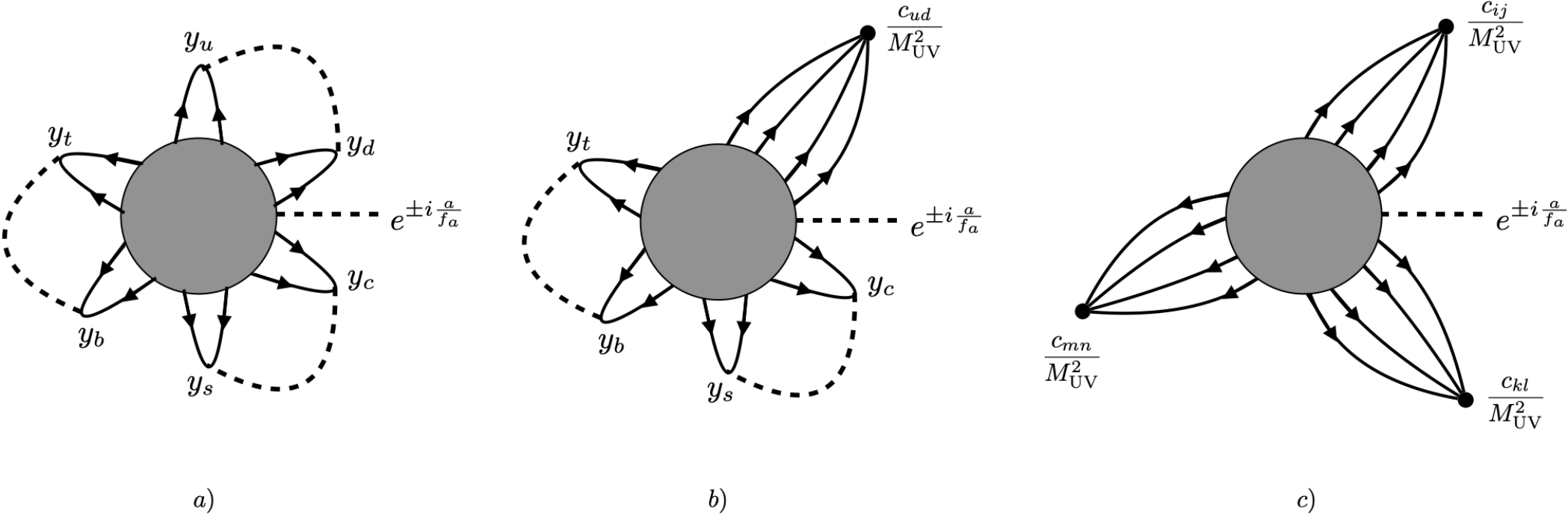}
	\caption{Six flavor QCD 't~Hooft operator closed up with Yukawa couplings a), Yukawa couplings and effective operators b) or only effective four-fermion operators c).}
	\label{fig:QCD6Higgs}
\end{figure}

%
As we have seen in Section~\ref{sec:EnhancedSmallInstantons} one of the simplest ways to enhance the contribution of small instantons to the axion potential is to add new colored matter which lowers the QCD beta function (see e.g.~\cite{Flynn:1987rs}). For concreteness we consider here ordinary QCD, which for six flavors has a beta function of $b^{{\rm QCD }_6}_0 = 7$, and assume for simplicity that at a scale $M$ above the top mass threshold a set of colored scalars is added to the theory which modifies the beta function to $b_0 < b^{{\rm QCD }_6}_0$.\footnote{If the additional particles are colored fermions, their contribution to the axion potential will be suppressed by their mass, which has to be inserted to close the zero modes. However, this mass suppression can be avoided if the spectrum also contains scalars with Yukawa couplings to the new fermions, which can be used to close up the zero modes.} Using only SM couplings the largest contribution to the axion potential originates from 't Hooft vertices closed with Higgs loops as shown in Figure~\ref{fig:QCD6Higgs}~a).
Extending the estimate of Eq.~\eqref{eq:InstantonHiggsUV} to six flavors we find that the contribution to the axion potential scales as
\be \label{eq:BetaFuncQCDAlign}
\delta V(a) \simeq -2 C_3  \left(\frac{8 \pi^2}{g^2 (M_{\rm UV})}\right)^{6} \left(\prod_i \frac{y_i}{(4\pi)}\right)  
    \left( \frac{M_{\rm UV}}{M} \right)^{4-b_0}\left(\frac{\Lambda_{{\rm QCD}_6}}{M}\right)^{b_0^{{\rm QCD}_6} -4}\Lambda_{{\rm QCD}_6}^4  \cos\left(\frac{a}{f_a}\right) \,, 
\ee
where $\Lambda_{{\rm QCD}_6}$ is the RG invariant scale of six flavor QCD and we assumed that $b_0 < 4$ which is required to get an enhancement of small instantons (see Eq.~\eqref{eq:InstantonHiggsUV}). We also assume that the gauge coupling coincides with the QCD coupling at the $Z$-pole and determine $g(M_{\rm UV})$ through RG running, taking into account mass thresholds. If the new particles have masses around the TeV scale, i.e. if we take $M = 1$~TeV we get up to $\mathcal{O}(1)$ factors
\be \label{eq:QCDBetaUV}
\delta V(a) \simeq - 10^{-28} \left( \frac{M_{\rm UV}}{1\text{ TeV}} \right)^{4-b_0} \Lambda_{{\rm QCD}_6}^4  \cos\left(\frac{a}{f_a}\right)\,.
\ee
For $M_{\rm UV}=10^{10}, 10^{12},10^{16}$~GeV this gives an $\mathcal{O}(1)$ contribution to the axion potential for $b_0 < 1.2, 1.67, 2.25$, respectively. However, as we will show momentarily, if the theory contains CP violating effective operators there will also be a misaligned contribution to the axion potential of similar size as Eq.~\eqref{eq:QCDBetaUV} or even larger. 

In order to be more concrete, assume there are operators of the form $\tfrac{c_{ij}}{M_{\rm UV}^2} \bar{\psi}_i \psi_i \bar{\psi}_j \psi_j + {\rm h.c.}$, where $c_{ij} = |c_{ij}| e^{i\delta_{ij}}$. For the suppression scale of the effective operator we take the UV-cutoff of the instanton calculation which can be interpreted as the scale of new physics. For $b_0 < 6$ any instanton contribution which uses effective operators to close the 't Hooft operator is UV dominated, such that diagrams with one insertion of the effective operator, as shown in Figure~\ref{fig:QCD6Higgs}~b), scale in the same way as diagrams with two or three insertions of an effective operator (see Figure~\ref{fig:QCD6Higgs}~c)). The only difference is the number of Yukawa couplings needed to close the remaining legs. For $|c_{ij}|\sim \mathcal{O}(1)$ it can be beneficial to use effective operators to close zero modes in order to avoid the Yukawa suppression. For $b_0 < 4$, and three insertions of the effective operator, the contribution to the axion potential scales as (see Eq.~\eqref{eq:InstEffOp})
\be\label{eq:SU3CPPotential}
\begin{split}
\delta V(a) \simeq -2 C_3  \left(\frac{8 \pi^2}{g^2 (M_{\rm UV})}\right)^6 &\frac{|c_{ij}| |c_{kl}| |c_{mn}|}{(4\pi)^6}  
    \left( \frac{M_{\rm UV}}{M} \right)^{4-b_0}\left(\frac{\Lambda_{{\rm QCD}_6}}{M}\right)^{b_0^{{\rm QCD}_6} -4}\\
&\times\Lambda_{{\rm QCD}_6}^4  \cos\left(\frac{a}{f_a} + \delta_{ij} +\delta_{kl}+\delta_{mn}\right)\,,
\end{split}
\ee
which for $|c_{ij}| \sim \mathcal{O}(1)$ is much larger than the contribution aligned with QCD from Eq.~\eqref{eq:BetaFuncQCDAlign} since one avoids the Yukawa suppression $\prod_i y_i \ll 1$. This implies that if one wants to raise the mass of the QCD axion purely by modifying the QCD beta function in the UV, i.e. the contribution to the axion mass from Eq.~\eqref{eq:QCDBetaUV} is of the same order or larger than from low-energy QCD, any CP violation not aligned with QCD will in general spoil the axion solution to the strong CP problem. It is legitimate to wonder about flavor and EDM constraints on the new CP-violating operators that we are introducing. They have $\mathcal{O}(1)$ phases and low energy probes of approximate SM symmetries are very sensitive to the effects that they induce. We do not discuss these constraitnts here because they decouple when $M_{\rm UV}\to \infty$, while the misaligned instanton contribution to the axion potential does not decouple if $b_0$ is sufficiently small, so there are always values of $M_{\rm UV}$ and $b_0$ for which instantons give the dominant effect.

\begin{figure}[t]
	\centering
	\subfigure{\includegraphics[width=0.48\textwidth]{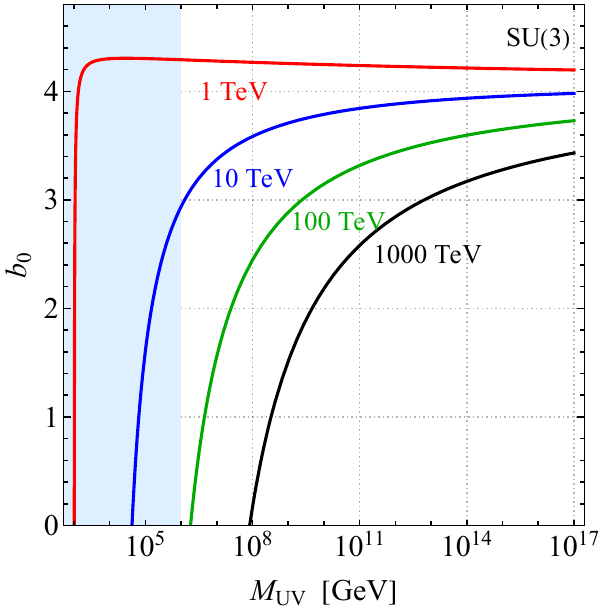}}\hfill
	\subfigure{\includegraphics[width=0.485\textwidth]{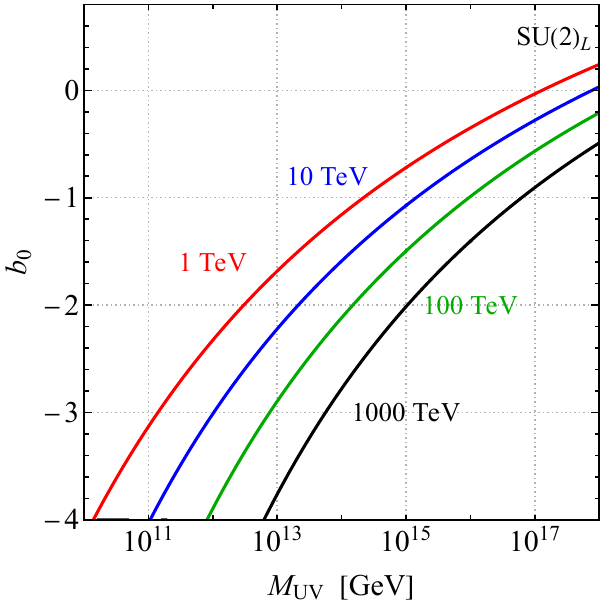}}
	\caption{Minimal value of $b_0$ for $SU(3)$ (left) and $SU(2)$ (right) as a function of the UV cutoff $M_{\rm UV}$ for which the induced $\bar{\theta}$ angle due to CP violating effective operators is $\bar{\theta}_{\rm ind} < 10^{-10}$. The different colors stand for various values of $M$, the scale where new charged particles appear. The blue shaded region shows the bound on $M_{\rm UV}$ from the contribution of CP violating higher-dimensional operators within low-energy QCD as discussed in Section~\ref{sec:QCDIR}. Note that we do not show further flavor and EDM bounds on $M_{\rm UV}$.}
	\label{fig:CPBounds}
\end{figure}

Even if the contribution to the axion mass from small instantons is negligible compared to the one from low-energy QCD it can still misalign the minimum of the axion potential. This can be phrased as a bound on the number of colored particles in the UV or equivalently the beta function coefficient $b_0$ which has to be satisfied in order for the axion solution to the strong CP problem not to be endangered. In the left panel of Figure~\ref{fig:CPBounds} we show the minimal value of $b_0$ as a function of the UV cutoff $M_{\rm UV}$ for several choices of $M$, where the new particles appear. In the plot we assumed $\mathcal{O}(1)$ Wilson coefficients and CP violating phases.
The region shaded in blue depicts the bound on the EFT cutoff from IR dominated contributions from effective CP violating operators in low-energy QCD which will be discussed in Section~\ref{sec:QCDIR}. 
Note that reaching $b_0 \sim 4$ requires the addition of $18$ scalars in the fundamental representation. Lower values of $b_0$ require even more colored particles.

Let us also mention that instead of adding additional colored matter, the beta function is also modified if QCD propagates in a flat extra dimension as discussed in~\cite{Gherghetta:2020keg}. The KK modes lead to an effective linear running of the gauge coupling and render the instanton calculation UV dominated. If the Wilson coefficient of CP violating operators are not tuned or small because of a symmetry their effect does not decouple, making an enhancement of the axion mass in the presence of CP violation impossible. This was also pointed out in~\cite{Bedi:2022qrd}.

%
\subsubsection{Misaligned axion potential from SU(2) instantons} \label{sec:SU2Instanton}
%
For general axion models $U(1)_{\rm PQ}$ is typically not only anomalous under $SU(3)_{\rm QCD}$ but also under $SU(2)_L$.
This implies that there will be a coupling of the form
\begin{equation}
	\left(\frac{a}{f_a} + \theta_{\rm EW}\right)\frac{g_w^2}{32\pi^2} W^A_{\mu\nu} \tilde{W}^{A\, \mu\nu}\,, 
 \label{eq:SU2L}
\end{equation}
where we neglected a possible non-trivial anomaly coefficient and $W^A_{\mu\nu}$ is the electroweak field strength and $\tilde{W}^A_{\mu\nu}$ its dual. In the SM, $\theta_{\rm EW}$ is unobservable since $U(1)_{B+L}$ is only broken by $SU(2)_L$~\cite{FileviezPerez:2014xju,Shifman:2017lkj} and therefore a $U(1)_{B+L}$ rotation can remove Eq.~\eqref{eq:SU2L} from the Lagrangian with no other effect on the SM. Therefore there is no contribution to the axion potential from $SU(2)_L$ instantons. 

However, if there is an additional explicit breaking of $U(1)_{B+L}$, e.g. through higher-dimensional operators, $SU(2)_L$ instantons contribute to the axion potential. If this is the case the axion quality problem might become an issue if $\bar{\theta}_{\rm EW} \neq \bar{\theta}_{\rm QCD}$.

Examples of higher-dimensional operators which violate $B+L$ are
\begin{equation}\label{eq:BLviolation}
	\frac{c_{L}^{ijkl}}{M_{L}^2} q^i q^j q^k L^l +\frac{c_R^{ijkl}}{M_R^2} (u^c)^i (u^c)^j (d^c)^k (e^c)^l + \ldots\,,
\end{equation}
where $i,j,k,l$ are generation indices. Such operators generically appear in the low-energy EFT of GUTs after the heavy (triplet) Higgses are integrated out.

\begin{figure}[t]
	\centering
	\includegraphics[width=\textwidth]{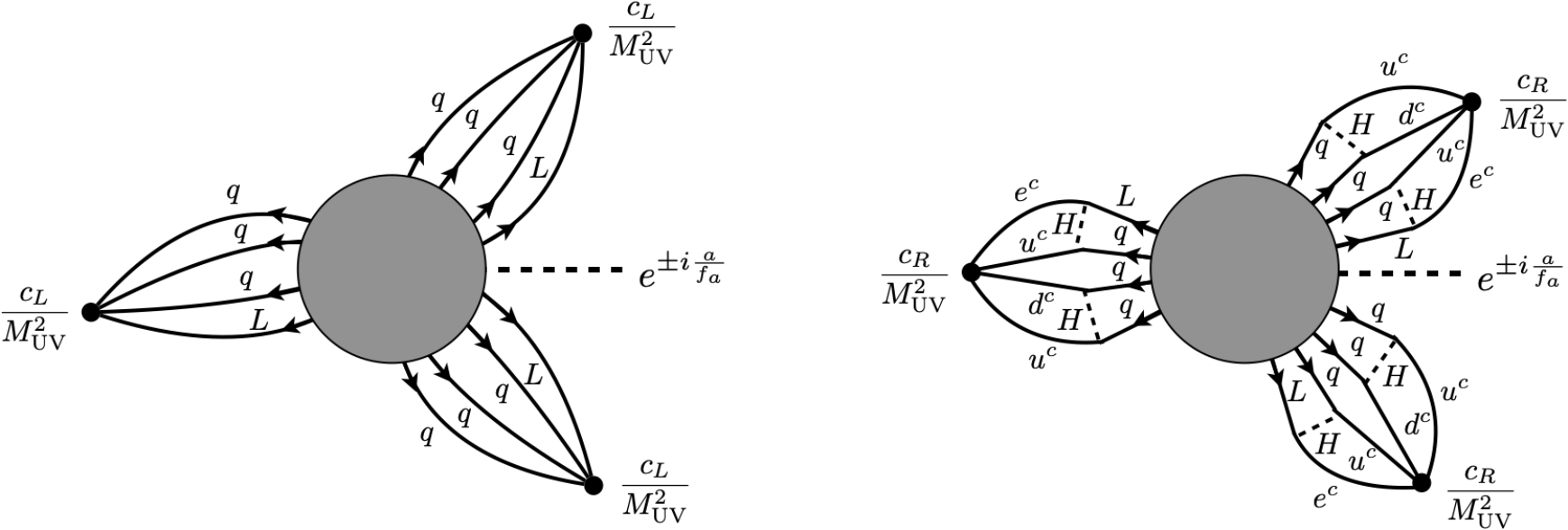}
	\caption{$SU(2)_L$ 't~Hooft operator closed up with the higher-dimensional operators in Eq.~\eqref{eq:BLviolation}.}
	\label{fig:tHooftOperatorSU2}
\end{figure}

Examples of contributions to the axion potential from insertions of these two effective operators are shown in Figure~\ref{fig:tHooftOperatorSU2}.
Without additional matter charged under $SU(2)_L$ the contribution from both diagrams is negligible. Note that the second diagram is additionally parametrically suppressed by $(4\pi)^{-12}(y_u y_c y_t)^2 (y_d y_s y_b) (y_e y_\mu y_\tau)\simeq 10^{-48}$ with respect to the first diagram which is why we will not consider it any further in the following. The contribution to the axion potential generated from the first diagram scales as
\begin{equation}\label{eq:Lambdasu2}
	\delta V_{SU(2)} (a) \simeq -2 C_2  \left(\frac{8 \pi^2}{g_w^2 (M_{\rm UV})}\right)^{4} \frac{|c_L|^3}{(4\pi)^6} \left( \frac{\Lambda_{SU(2)}}{M_{\rm UV}}\right)^{b_0^{SU(2)} -4} \Lambda_{SU(2)}^4 \cos\left(\frac{a}{f_a} + \theta_{\rm EW} + 3\delta_{c_L}\right)\,,
\end{equation}
where $\Lambda_{SU(2)}$ is the RG invariant scale of $SU(2)$ and $\delta_{c_L}$ the phase of $c_L$ which for simplicity we assume to be universal for all generations. For $b_0^{\rm SU(2)}<10$, which is always the case for $SU(2)$ gauge theories, the integral over the instanton size is UV dominated and cut off at $1/M_{\rm UV}$ which we identify with the EFT cutoff and suppression scale of the effective operator. Note that $SU(2)_L$ is spontaneously broken in the IR, such that instantons of size $\rho \gg 1/(g_w v)$ are exponentially suppressed.

In the SM $b_0^{SU(2)} = \tfrac{19}{6}$ with an RG invariant scale of $\Lambda_{SU(2)}\simeq 3\cdot 10^{-24}$~GeV, which makes the contribution numerically insignificant 
\begin{equation}\label{eq:Lambdasu2SM}
	\frac{\delta V_{SU(2)} (a)}{\cos\left(\frac{a}{f_a} + \theta_{\rm EW} + 3\delta_{c_L}\right)} \simeq -\left( 2\cdot 10^{-15}\text{ GeV} \right)^4\left(\frac{g_w^2(10^{16}\text{ GeV})}{g_w^2(M_{\rm UV})}\right)^4 \left( \frac{M_{\rm UV}}{10^{16}\text{ GeV}}\right)^{5/6} \,.
\end{equation}
However, since the contribution to the potential is dominated by small instantons, this is a UV dependent statement. UV modifications of $SU(2)_L$, such as the non-trivial embedding into a larger gauge group (see e.g.~\cite{Morrissey:2005uza,Fuentes-Martin:2014fxa}) or additional matter charged under $SU(2)_L$, can further enhance small instantons. As an example let us assume as before that at a scale $M$ a set of scalars charged under $SU(2)_L$ is added to the spectrum, modifying the beta function coefficient to $b_0 < b_0^{SU(2)}$. Incorporating the scale matching condition at $M$ leads to an expression analogous to Eq.~\eqref{eq:SU3CPPotential}
\be
\begin{split}
\delta V_{SU(2)}(a) \simeq -2 C_2  \left(\frac{8 \pi^2}{g_w^2(M_{\rm UV})}\right)^{4} &\frac{|c_{L}|^3}{(4\pi)^6}  
    \left( \frac{M_{\rm UV}}{M} \right)^{4-b_0}\left(\frac{\Lambda_{SU(2)}}{M}\right)^{b_0^{SU(2)} -4}\\
&\times\Lambda_{SU(2)}^4  \cos\left(\frac{a}{f_a} + \theta_{\rm EW} +3 \delta_{c_L}\right)\,.
\end{split}
\ee
Assuming $\mathcal{O}(1)$ Wilson coefficients we can again find the minimal value of $b_0$ as a function of the UV cutoff $M_{\rm UV}$ for which $\bar{\theta}_{\rm ind} < 10^{-10}$. As shown in the right panel of Figure~\ref{fig:CPBounds} 
the beta function coefficient in general has to switch sign in order for the axion solution to the strong CP problem to be in danger.

Let us finally mention that instanton contributions to the mass of electroweak axions have also been discussed in~\cite{McLerran:2012mm,Nomura:2000yk,Ibe:2018ffn} in the context of axionic quintessence models and in~\cite{Dine:2022mjw} mainly in the context of Grand Unified Theories.
%
\subsection{IR Dominated Misaligned Contributions in QCD}\label{sec:QCDIR}
%
In this work we are mainly interested in UV instanton effects on the magnitude and alignment of the axion potential. However, for completeness, in this section we review very briefly the impact that a CP-violating operator, generated at high-energy, can have on this potential, purely from the IR dynamics of QCD. These effects are also present when UV instantons  are negligible.

Assuming there is no non-trivial UV modification of QCD 
the SM QCD contribution to the axion potential is dominated by non-perturbative effects in the IR at the QCD scale. This contribution cannot be reliably computed with instantons as QCD gets strongly coupled and all order instanton contributions are equally important. In fact instantons do not even give the leading contribution to the axion potential (see e.g.~\cite{Csaki:2023yas} for a recent discussion). CP violating effective operators, such as $\tfrac{|c_{ud}| }{M_{\rm UV}^2}e^{i\delta_{ud}} \bar{u} u \bar{d} d + {\rm h.c.}$ nonetheless affect the axion potential and generate a misaligned contribution.\footnote{Note that some effective operators can also directly contribute to the neutron dipole moment~\cite{Bigi:1990kz,Pospelov:2000bw,Pospelov:2005pr,Choi:2023bou}.} It is tempting to estimate this effect by using the effective operator to close up zero modes in the 't Hooft operator, however in the strongly-coupled regime the instanton calculation is IR divergent and does not provide a reliable estimate. Thus in order to estimate the effect one has to resort to non-perturbative methods. In order to do so it is more convenient to parameterize the axion potential in terms of the topological susceptibility $\chi (0)$ and an additional correlator $\chi_{\mathcal{O}}(0)$ which parameterizes the misalignment due to an effective operator $\mathcal{O}$. Up to quadratic order in the axion field its potential can be expressed as
\be \label{eq:AxionPotChi}
V(a) = \chi_{\mathcal{O}}(0) \frac{a}{f_a} + \frac{1}{2} \chi (0) \left(\frac{a}{f_a}\right)^2+\ldots\,.
\ee
Both $\chi_{\mathcal{O}}(0)$ and $\chi (0)$ can be expressed in terms of correlators of the form~\cite{Shifman:1979if,Bigi:1990kz,Witten:1979vv,Pospelov:2005pr}
\begin{align}
    \chi(0)&=-i \lim _{k \rightarrow 0} \int d^4 x e^{i k x}\left\langle 0\left|T\left\{\frac{1}{32 \pi^2} G^a_{\mu\nu}\tilde{G}^{a\,\mu\nu}(x) \frac{1}{32 \pi^2} G^a_{\mu\nu}\tilde{G}^{a\,\mu\nu}(0)\right\}\right| 0\right\rangle\,,\\
    \chi_{\mathcal{O}}(0) &= -i \lim _{k \rightarrow 0} \int d^4 x e^{i k x}\left\langle 0\left|T\left\{\frac{g^2}{32 \pi^2} G^a_{\mu\nu}\tilde{G}^{a\,\mu\nu}(x) \frac{c_{\mathcal{O}}}{M_{\rm UV}^2} \mathcal{O}(0)\right\}\right| 0\right\rangle\,,
\end{align}
which have to be evaluated using non-perturbative methods, such as QCD sum rules (see e.g.~\cite{Crewther:1977ce,Shifman:1979if,Pospelov:1999rg}). The explicit definition of $\chi_{\mathcal{O}}(0)$ and $\chi (0)$ and the way they enter the axion potential in Eq.~\eqref{eq:AxionPotChi} can be intuitively understood by noticing that in the QCD Lagrangian one can choose a basis in which $\tfrac{a}{f_a} G_{\mu\nu}^a \tilde{G}^{a\, \mu\nu}$ is the only non-derivative axion coupling. Thus when computing contributions to the potential each power of the axion field has to be accompanied with a $G_{\mu\nu}^a \tilde{G}^{a\, \mu\nu}$ factor in a vacuum-to-vacuum correlator. 
 A simple estimate using dimensional analysis assuming that only one power of the effective operator enters the computation yields an induced $\theta$ angle of
\be \label{eq:CPlowEQCD}
\bar{\theta}_{\rm ind} \simeq c_{\mathcal{O}} \frac{\Lambda_{\rm QCD}^2}{M_{\rm UV}^2} \, \sin (\delta) \sim c_{\mathcal{O}}\frac{m_\pi f_\pi}{M_{\rm UV}^2}\, \sin (\delta) \quad \Rightarrow\quad  \frac{M_{\rm UV}}{c_{\mathcal{O}}^{1/2}} \gtrsim   10^6\text{ GeV}\cdot \sin^{1/2}(\delta)\,,
\ee
where  $c_{\mathcal{O}}$ and $\delta$ are the magnitude and phase of the Wilson coefficient of the effective operator and we used the experimental bound $\bar{\theta} < 10^{-10}$. In this estimate we used that the contribution from one insertion of the effective operator is proportional to $c_{\mathcal{O}}\sin (\delta)/M_{\rm UV}^2$ and that $\Lambda_{\rm QCD} \simeq (m_\pi f_\pi)^{1/2}$ is the only other relevant dimensionful scale. This estimate is consistent with the findings of~\cite{Bedi:2022qrd}. The important difference with respect to UV-instantons is that these effects decouple as $M_{\rm UV}\to \infty$, even if we do arbitrary violence to $b_0$ in the UV.
%
\subsection{Misaligned Axion Potential}
%
We have seen in the previous examples that small instanton corrections to the axion potential may or may not be aligned with the QCD contribution. Aligned contributions simply raise the axion mass and can lead to interesting novel phenomenology for the axion. Misaligned contributions however usually exacerbate the axion quality problem by leading to additional corrections to $\bar\theta$, and hence should ideally be avoided. We can parameterize these new contributions to the axion potential 
using the general form 
\begin{equation} \label{eq:axionPot}
	-V (a) = \left( m_\pi^2 f_\pi^2 + \Lambda_{\rm SI}^4 \right) \cos\left(\frac{a}{f_a}\right) + \Lambda_{\rm CPV}^4 \cos\left(\frac{a}{f_a}+\delta\right)\,,
\end{equation}
where $m_\pi^2 f_\pi^2$ is the contribution from low-energy QCD, $\Lambda_{\rm SI}^4$ is a possible aligned contribution from small instantons (SI stands for small instantons) and $\Lambda_{\rm CPV}^4$ is the misaligned contribution generated from CP violating sources with an additional phase $\delta$ in the axion potential. Of course if there is no additional source of CP violation 
present in the UV theory, then   instantons will not contribute to $\Lambda_{\rm CPV}$. 

In one of the examples above, Eq.~\eqref{eq:BetaFuncQCDAlign}, the 't~Hooft operator closed with Higgs loops is aligned with the low-energy QCD axion potential and therefore contributes to $\Lambda_{\rm SI}^4$. However, closing the 't~Hooft operator with higher-dimensional operators as in Eq.~\eqref{eq:SU3CPPotential} contributes to $\Lambda_{\rm CPV}^4$ if the Wilson coefficient has a non-vanishing CP violating phase. In the limit $\Lambda_{\rm CPV}^4 \ll (m_\pi^2 f_\pi^2 + \Lambda_{\rm SI}^4)$ this induces an axion VEV or effective $\bar{\theta}$ angle
\be\label{eq:axionVEV}
\bar{\theta}_{\rm ind} \equiv \left\langle \frac{a}{f_a}\right\rangle = \frac{\Lambda_{\rm CPV}^4}{f_\pi^2 m_\pi^2 + \Lambda_{\rm SI}^4} \sin(\delta) < 10^{-10}\,.
\ee
$\Lambda_{\rm CPV}^4$ and $\Lambda_{\rm SI}^4$, if both present, are often related, since enhanced small instantons typically also enhance the effects of CP violating operators. However, this is not necessarily the case as we saw in Section~\ref{sec:SU2Instanton} in the example where $SU(2)_L$ instantons only generate $\Lambda_{\rm CPV}^4$. 

If both $\Lambda_{\rm CPV}^4$ and $\Lambda_{\rm SI}^4$ are present we can distinguish between two different scenarios.

\paragraph{CP violating effects decouple ($\Lambda_{\rm CPV}/ \max[\Lambda_{\rm SI}, m_\pi^2f_\pi^2] \rightarrow 0$ for $M_{\rm UV}\rightarrow\infty$):}

Here we assume a common cutoff $M_{\rm UV}$, for the integral over instanton sizes, and the suppression scale of the higher-dimensional operator, e.g. $\tfrac{c_{ij}}{M_{\rm UV}^2} \bar{\psi}_i \psi_i \bar{\psi}_j \psi_j + h.c.$. 
If $\Lambda_{\rm CPV}/ \Lambda_{\rm SI} \rightarrow 0$ for $M_{\rm UV}\rightarrow\infty$, the CP violating effects decouple. This can only happen if 
$\Lambda_{\rm SI}$ is generated 
at a scale $M_{\rm SI} \ll M_{\rm UV}$ much below the suppression scale of the effective operator and UV cutoff. This naturally occurs if the contribution to $\Lambda_{\rm SI}$ is IR dominated and the one to $\Lambda_{\rm CPV}$ is UV dominated. In this case $\bar{\theta}_{\rm ind} \propto \Lambda_{\rm CPV}^4 / \Lambda_{\rm SI}^4 \propto M_{\rm SI}^2 / M_{\rm UV}^2$ for the insertion of one dimension-six operator. This never happens in our two simple examples discussed in Section~\ref{sec:SmallInstanton}, where $\Lambda_{\rm SI}$ and $\Lambda_{\rm CPV}$ are dominated by the same instanton sizes, regardless of the $\beta$-function.  However, this behavior can be observed in many UV modifications of QCD that were studied in~\cite{Bedi:2022qrd}. These include non-trivial embeddings of QCD into a larger gauge group~\cite{Agrawal:2017ksf,Agrawal:2017evu,Csaki:2019vte} and mirror QCD~\cite{Berezhiani:2000gh,Hook:2014cda,Rubakov:1997vp,Fukuda:2015ana,Hook:2019qoh}. By performing the explicit computation the authors found that one has to require that $M_{\rm UV} \gtrsim (10^5 - 10^8) M_{\rm SI}$ in order not to spoil the axion solution to the strong CP problem~\cite{Bedi:2022qrd}. 

\paragraph{CP violating effects do not decouple ($\Lambda_{\rm CPV} \sim \Lambda_{\rm SI}$):}
If we again identify the UV cutoff of the EFT and instanton size integral with the suppression scale of the higher-dimensional operator, CP violating effects do not decouple if both $\Lambda_{\rm SI}$ and $\Lambda_{\rm CPV}$ are generated by UV dominated instantons. In this case we naturally find $\Lambda_{\rm CPV}^4 \sim \Lambda_{\rm SI}^4$. If these are comparable to the low-energy QCD contribution to the axion potential this implies that an $\mathcal{O}(1)$ CP violating phase $\delta$ prevents a successful axion solution to the strong CP problem if the Wilson coefficient of the effective operator is not tuned or protected by symmetries. An example of such a setup is the modified QCD beta function in Section~\ref{sec:QCDBetaFunction} or 5D small instantons~\cite{Gherghetta:2020keg}.
%
%
\section{UV-safe models: Composite Axions and $Z_N$ Axions}\label{sec:UVSafe}
%
As we have demonstrated above, even if the axion is equipped with a high-quality Peccei-Quinn symmetry, a successful axionic solution to the strong CP problem crucially depends on UV physics. If the UV theory contains CP violating couplings, which is in general expected, any modification of the particle spectrum or the running of the gauge coupling can endanger the solution to the strong CP problem. For this reason it would be desirable to identify models which do not suffer from this UV sensitivity.

In this section we want to present two models which have this property: $Z_N$ axions~\cite{Hook:2018jle} and composite axions~\cite{Contino:2021ayn}. They achieve the UV insensitivity in different ways. $Z_N$ axions possess a discrete shift symmetry which protects them from $m$-instanton effects with $m < N$, implying that as long as the coupling  is perturbative, the usual  exponential suppression of small instantons is enhanced. 

Composite axions~\cite{Contino:2021ayn}, on the other hand, do not exist as elementary degrees of freedom above the compositeness scale. Misalignment effects can therefore only originate from 
low-energy PQ-breaking. If the model does not allow such operators up to a high mass dimension, such as in~\cite{Contino:2021ayn}, misaligned contributions to the axion potential from UV instantons are irrelevant. Note however, that this does not prevent misalignment effects in low-energy QCD along the lines of Section~\ref{sec:QCDIR}. These, however, decouple as $M_{\rm UV}\to \infty$.
%
\subsection{$Z_N$ Axions}
%
A class of UV-safe models are axions in a $Z_N$ symmetric world. Proposed by Hook~\cite{Hook:2018jle}, and further investigated in~\cite{DiLuzio:2021pxd}, such models assume that there are $N$ identical copies of the SM related by a $Z_N$ symmetry, but a single axion coupling to all the sectors. Under the $Z_N$ symmetry we have
\begin{align}
	{\rm SM}_k &\xrightarrow{Z_{N}} {\rm SM}_{k+1}\,,\\
	\frac{a}{f_a} &\xrightarrow{Z_{N}} \frac{a}{f_a} + \frac{2\pi}{N}\,,
\end{align}
where $k=0,\ldots, N-1$ and SM$_{N} =$ SM$_{0}$. As we will see momentarily for odd $N$ the strong-CP problem will be solved in the $k=0$ sector, which we identify with the sector that we live in. The $Z_N$-symmetric axion couplings to the different sectors are given by
\begin{equation}
	\mathcal{L}_a = \sum_{k=0}^{N-1} \left(\frac{a}{f_a} + \frac{2\pi k}{N} + \bar{\theta}\right) \frac{g^2}{32\pi^2} G_{k\, \mu\nu}^a \tilde{G}_k^{a\,\mu\nu}\,.
\end{equation}
The axion potential from low-energy QCD can be computed within chiral perturbation theory. Summing up the contributions from all sectors in 2-flavor chiral perturbation theory yields~\cite{Hook:2018jle}\footnote{Note that this is equivalent to our Eq.~\eqref{eq:VIR}.}
\begin{equation}
	V_{N} (a) = -m_\pi^2 f_\pi^2  \frac{m_d}{m_u + m_d} \sum_{k=0}^{N-1} \sqrt{1 + z^2 + 2 z\cos\left( \frac{a}{f_a}+\frac{2 \pi k}{N}\right)}\,,
\end{equation}
where $z=m_u / m_d \approx 1/2$ and we have absorbed $\bar{\theta}$ into the axion field. Expanding the above in powers of $z$ one can see that the first contribution to the axion potential arises at order $z^{N}$. All lower-order terms cancel or are independent of the axion field due to a set of trigonometric identities which for $N > m \geq 0$ can be written as~\cite{Hook:2018jle}
\begin{equation}
	\sum_{k=0}^{N-1} \cos ^m\left(\frac{a}{f_a}+\frac{2 \pi k}{N}\right)= \begin{cases}0 & m=\text { odd } \\ \frac{N}{2^m} \frac{m !}{(m / 2) ! !^2} & m=\text { even }\end{cases}\,,
\end{equation}
Thus the first axion-dependent contribution arises at $m=N$, leading to an exponential suppression of the axion mass. In the large $N$ limit the potential has the form~\cite{DiLuzio:2021pxd}
\begin{equation}\label{eq:ZNpotential}
V_{N}\left(a\right) \simeq \frac{m_\pi^2 f_\pi^2}{\sqrt{\pi}} \sqrt{\frac{1-z}{1+z}} N^{-1 / 2}(-1)^{N} z^{N} \cos \left( \frac{N a}{f_a}\right)\,,
\end{equation}
such that $\langle a \rangle = 0$ is a minimum for odd $N$. This implies that for odd $N$ the strong CP problem is still solved in one of the sectors, i.e. the $k=0$ sector, what corresponds to a tuning of the order of $1/N$, from the discrete choice of living at $k=0$ to solve the problem in our sector. Also note that the axion mass from Eq.~\eqref{eq:ZNpotential} is exponentially suppressed and scales as $m_a^2 f_a^2 \propto m_\pi^2 f_\pi^2\, N^{3/2} z^{N}$.

Let us now consider small instanton contributions to the axion potential from all $N$ sectors, i.e. we restrict ourselves to an energy range where the coupling is perturbative. Due to the $Z_N$ symmetry the one-instanton contributions from all sectors are identical except for a different effective $\bar{\theta}_k$ angle of $\bar{\theta}_k = \bar{\theta} + \tfrac{2\pi k}{N}$, which results in different phases for instanton contributions from different sectors. The different phases from the $N$ sectors lead to a cancellation of the whole potential
\begin{equation}
	V_a = - \Lambda_{\rm SI}^4 \sum_{k=0}^{N-1} \cos\left(\frac{a}{f_a} + \frac{2\pi k}{N} \right) =0\,,
\end{equation}
where we again absorbed $\bar{\theta}$ in the axion field and introduced the dimensionful scale $\Lambda_{\rm SI}$ generated by small instantons in each sector. 

Thus any small one-instanton contribution in the $Z_N$ model does not give a contribution to the axion potential. The first non-vanishing contribution originates from an $N$-instanton effect which is suppressed by $\exp (-N \tfrac{8\pi^2}{g^2})$. The above considerations also apply to contributions from CP violating effective operators which must have the same phase in all sectors due to the $Z_N$ symmetry.

Hence the $Z_N$ axion is by construction safe from CP violating small instanton contributions. 
However, if the $Z_N$ symmetry is explicitly broken this is not necessarily true anymore. 
Furthermore the reduced axion mass in the $Z_N$ model (see Eq.~\eqref{eq:ZNpotential}) leads to a more severe misalignment than in ordinary axion models if the $Z_N$ symmetry is broken. 
As an example of this effect let us consider the model introduced in Ref.~\cite{Banerjee:2022wzk} where the axion also plays the role of the relaxion. Note that for the rest of this section we will slightly modify our notation to make contact with~\cite{Banerjee:2022wzk}. Our previous identification $\Lambda_{\rm QCD}^4 \simeq m_\pi^2 f_\pi^2$ assumed a Higgs VEV of $v = \langle H \rangle =174$~GeV. However, in a relaxion setup the Higgs VEV is not fixed and $m_\pi^2 f_\pi^2$ scales approximately\footnote{In~\cite{Banerjee:2022wzk} they considered a linear scaling, neglecting the dependence of the running of $\alpha_s$ on $v$ due to quark thresholds. If we include this effect, $f_\pi\sim v^{0.3}$ for $v_{\rm SM}\lesssim v \lesssim 10^4 v_{\rm SM}$. Note that when $v \gtrsim 10^4 v_{\rm SM}$ up and down quarks become heavier than the QCD scale and also the dependence of $m_\pi^2$ on $v$ changes. Our modified $v^2$ scaling does not affect the conclusions of~\cite{Banerjee:2022wzk}.} as $v^2$. In order to make this more explicit we make the replacement $m_\pi^2 f_\pi^2 = \tfrac{1+z}{z} y_u v^2\, \tilde{\Lambda}_{\rm QCD}^2$ where $z$ and $\tilde{\Lambda}_{\rm QCD}$ are approximately independent of the Higgs VEV. Note that  $\tilde{\Lambda}_{\rm QCD}\equiv (z m_\pi^2 f_\pi^2/(1+z)y_u v^2)_{\rm SM}$ is smaller than the usual $\Lambda_{\rm QCD}$. In terms of this new parameter the relaxion-Higgs potential considered in~\cite{Banerjee:2022wzk} takes the form
\begin{equation}
\begin{split}
	&V(a,H) = (M_{\rm UV}^2 - g M_{\rm UV} a) |H|^2 + \lambda |H|^4 + V_{\rm roll}(a) + V_{\rm br}(a ,\langle H\rangle )\,,\\
	&V_{\rm roll} (a) = - g M_{\rm UV}^3 a \,,\quad V_{\rm br}(a,\langle H\rangle ) = (-1)^N \tilde{\Lambda}_{\rm QCD}^2 y_u v^2 \kappa \cos \left( \frac{N a}{f_a}\right)\,,
\end{split}
\end{equation}
where the backreaction potential is the one of Eq.~\eqref{eq:ZNpotential} with $\kappa = z^{N-1}\sqrt{(1-z^2)/(\pi N)}$, $M_{\rm UV}$ is the UV cutoff and $g$ a dimensionless coupling constant. In this setup the relaxion stops when $V^\prime_{\rm roll} \simeq V_{\rm br}^\prime$ which happens for $N a/f_a \sim 3\pi/2$ for $N$ even and at $N a/f_a \sim \pi/2$ for $N$ odd~\cite{Banerjee:2020kww}. For both cases there is no SM sector in which the strong CP problem is solved.\footnote{The effective $\bar{\theta}$ angle in the $k$-th sector is $\bar{\theta}_{\rm eff}^{(k)} = \frac{\langle a\rangle}{f_a} + \frac{2\pi k}{N}$.} For this reason the authors introduce an explicit breaking of the $Z_N$ symmetry in the $k=0$ sector (that is no longer our sector in this model), where they take $y_u' > y_u^{\rm SM}$ and also $v' \geq v$, what also leads to $\tilde{\Lambda}_{\rm QCD}' \geq \tilde{\Lambda}_{\rm QCD}$. Due to this explicit breaking not all $z^1$ terms cancel exactly and one obtains a back-reaction potential of the form\footnote{Note that due to the breaking of the $Z_N$ symmetry, not only does the leading term in $z$ not cancel but also higher order terms. As was pointed out in~\cite{Banerjee:2022wzk} these play an important role in the dynamics but here we only consider the leading order term.}
\begin{equation}
V_{\mathrm{br}}(a) \sim-\tilde{\Lambda}_{\mathrm{QCD}}^{\prime 2} y_u^{\prime} (v^{\prime})^2 \cos \left(\frac{a}{f_a}\right)\left(1-\epsilon_b \gamma\right) -\tilde{\Lambda}_{\mathrm{QCD}}^2 y_u v^2 \kappa \cos \left(\frac{N a}{f_a}\right)\,,
\end{equation}
where $\epsilon_b = \tilde{\Lambda}_{\mathrm{QCD}}^2 y_u/(\tilde{\Lambda}_{\mathrm{QCD}}^{\prime 2} y_u^{\prime})$ and $\gamma = (v/v')^2$ parameterize the breaking of the $Z_N$ symmetry. The first term is simply the leading contribution in $z$ from the $k=0$ sector axion potential. For $\epsilon_b \gamma =1$ we recover the $Z_N$ symmetry and this term cancels. However, if the first term dominates, i.e. if $\epsilon_b \gamma \ll 1$ the relaxion stopping point is determined by this term and occurs at $a/f_a \sim \pi/2$. This implies that for $N=4 n$, where $n$ is a positive integer, there is one sector in which the strong CP problem is solved.

However, the larger QCD scale in the $k=0$ sector together with the small axion mass at the relaxion stopping point make the setup vulnerable to misaligned contributions from CP violating higher-dimensional operators. 
E.g. if we consider the insertion of either $f^{abc}G^a_{\mu\nu} G^{b\, \nu\rho} \tilde{G}^c_\rho\,^\mu$ (see e.g.~\cite{Bedi:2022qrd}) or a CP violating four fermion operator we expect to get an additional contribution to the potential of size
\begin{equation}\label{eq:ZNCPV}
	\delta V (a) = |c_{\delta'}| \frac{\tilde{\Lambda}_{\rm QCD}'^2}{M_{\rm UV}^2}\tilde{\Lambda}_{\rm QCD}^{\prime 2} y_u' v^{\prime 2} \left(1-\epsilon_b \gamma\right) \cos \left(\frac{a}{f_a} + \delta'\right)\,,
\end{equation}
where $c_{\delta'} = |c_{\delta'}| e^{i \delta'}$ is the Wilson coefficient of the operator and $\delta'$ is an $\mathcal{O}(1)$ C P-violating phase and $M_{\rm UV}$ the suppression scale of the CP-violating operator. Here we assumed one insertion of the CP-violating operator and used dimensional analysis to fix the dependence on $\tilde{\Lambda}_{QCD}'$. The $\left(1-\epsilon_b \gamma\right)$ factor arises from the fact that this contribution also vanishes in the $Z_N$ symmetric limit.

Let us now estimate for which values of $M_{\rm UV}$ this would misalign the axion potential and spoil the axion solution to the strong CP problem. Using Eq.~(III.22) of~\cite{Banerjee:2022wzk} we consider the axion potential around its stopping point with the addition of the misalignment contributions from CP violating operators in Eq.~\eqref{eq:ZNCPV}
\begin{equation}
	V(a) = |c_{\delta'}| \frac{\tilde{\Lambda}_{\rm QCD}'^2}{M_{\rm UV}^2}\tilde{\Lambda}_{\rm QCD}'^2 y_u' v'^2 \left(1-\epsilon_b \gamma\right) \cos \left(\frac{a}{f_a} + \delta'\right) - m_a^2 f_a^2\cos \left(\frac{a}{f_a}\right)\,,
\end{equation}
where $m_a^2 f_a^2 = \delta \tilde{\Lambda}_{\rm QCD}^2 y_u v^2/(\epsilon_b \gamma)$. Finding the minimum via
\begin{equation}
	0{=} V'(a) = |c_{\delta'}|\frac{\tilde{\Lambda}_{\rm QCD}'^2}{M_{\rm UV}^2}\tilde{\Lambda}_{QCD}'^2 y_u' v'^2 \sin(\delta') - \frac{\delta \tilde{\Lambda}_{\rm QCD}^2 y_u v^2}{\epsilon_b \gamma} \frac{a}{f_a} + \mathcal{O} (a^2)\,,
\end{equation} 
where we expanded in small $a$ and dropped subleading terms. Solving this for $a$ we find
\begin{equation}
	\left|\frac{a}{f_a}\right| = \left| |c_{\delta'}|\frac{\tilde{\Lambda}_{\rm QCD}'^2}{M_{\rm UV}^2} \frac{\sin(\delta')}{\delta} \right| \lesssim 10^{-10}\,.
\end{equation}
Following Eq.~(III.20) in~\cite{Banerjee:2022wzk} we take
\be
\delta \simeq 4 \times 10^{-11}\left(\frac{10^6 \mathrm{GeV}}{M_{\rm UV}}\right) \sqrt{\frac{\gamma}{\epsilon_b}}\, , 
\ee
and with $\sqrt{\gamma / \epsilon_b} \sim  |c_{\delta'}| \sin(\delta') \sim \mathcal{O}(1)$ we find
\begin{equation}
	M_{\rm UV} \geq 10^{10} \tilde{\Lambda}_{\rm QCD}' \sim 10^{13}\text{ GeV}\,,
\end{equation}
where we used the benchmark value $\tilde{\Lambda}_{\rm QCD}' \sim $~TeV from~\cite{Banerjee:2022wzk}.
This is a much stronger bound than in pure QCD, where one finds $M_{\rm UV} \gtrsim 10^{6}\text{ GeV}$ from similar considerations (see Eq.~\eqref{eq:CPlowEQCD}). Thus while Planck suppressed Peccei-Quinn breaking operators are not an issue in this model for large enough $N$ and small enough $f_a$, CP-violating dimension-6 operators with a suppression scale as high as $10^{13}$ GeV can be enough to spoil the solution to the strong CP problem. Note that generic explicit breaking of the $Z_N$ symmetry would also allow a new $\theta$-angle in the $k=0$ sector that would misalign the axion potential. Here we followed~\cite{Banerjee:2022wzk} and assumed a specific source of $Z_N$ breaking in the UV theory, namely $y_u' > y_u$ and $v' > v$ in the $k=0$ sector.
%
\subsection{Composite Axions}
%
Another setup that provides protection from misaligned small instantons are composite axion models with a high-quality PQ symmetry, as proposed in~\cite{Contino:2021ayn}. The axion is only a relevant degree of freedom below the confinement scale of a new gauge group and is therefore screened from UV contributions to its potential. This is interesting to us because these models typically contain a large number of particles charged under $SU(3)$ or $SU(2)_L$ which might enhance small instantons. In the following we outline the structure of these models with an emphasis on possible effects from small instantons.

The models in~\cite{Contino:2021ayn} assume a gauge symmetry $SU(N_{\rm DC})\times U(1)_D \times G_{\rm SM}$, with additional fermions which are vector like under $SU(N_{\rm DC})\times G_{\rm SM}$ but chiral under the full gauge group due to their $U(1)_D$ charge assignment, i.e. we have LH fermions $\psi_i$ and $\chi_i$ with quantum numbers
\begin{equation}
	\psi_i \sim\left(\square, p_i, r_i\right), \quad \chi_i \sim\left(\bar{\square}, q_i, \bar{r}_i\right), \quad i=1, \ldots, n_f
\end{equation}
under $SU(N_{\rm DC})\times U(1)_D \times G_{\rm SM}$, where the $U(1)_D$ charges are chosen such that no mass terms are allowed, in particular $p_i \neq - q_i$ for irreducible SM representations $r_i$ and $\bar{r}_i$. The BSM part of the Lagrangian is therefore given by
\begin{equation}
	\mathcal{L}_{\mathrm{BSM}}=-\frac{1}{4} \mathcal{G}_{\mu \nu}^a \mathcal{G}^{a, \mu \nu}-\frac{1}{4} F_{\mu \nu}^D F^{D \mu \nu}+\frac{\varepsilon}{2} F_{\mu \nu}^D B^{\mu \nu}+\sum_i \psi_i^{\dagger} i D_\mu \bar{\sigma}^\mu \psi_i+\sum_i \chi_i^{\dagger} i D_\mu \bar{\sigma}^\mu \chi_i\,,
\end{equation}
where $i$ sums over irreducible representations and $\epsilon$ parameterizes a possible kinetic mixing between $U(1)_D$ and $U(1)_Y$. 
Let us have a look at the minimal model to understand how the mechanism works.\footnote{Note that the minimal model does not have a high-quality PQ symmetry and allows for PQ violating operators already at dimension 6. However, if small instanton effects are irrelevant in this scenario they are even more so in safer models.} The particle content is given by
\begin{equation}
	\begin{array}{cccc} 
& \mathrm{SU}\left(N_{\mathrm{DC}}\right) & \mathrm{U}(1)_{\mathrm{D}} & \mathrm{G}_{\mathrm{SM}} \\
\hline \psi_1 & \square & +1 & 1 \\
\psi_2 & \square & -1 & 1 \\
\psi_3 & \square & +1 & r \\
\psi_4 & \square & -1 & \bar{r} \\
\hline \chi_1 & \bar{\square} & -q & 1 \\
\chi_2 & \bar{\square} & +q & 1 \\
\chi_3 & \bar{\square} & -q & \bar{r} \\
\chi_4 & \bar{\square} & +q & r
\end{array}
\end{equation}
with $r=(\mathbf{3},\mathbf{1})_y$ of $(SU(3),SU(2)_L)_{U(1)_Y}$ for an arbitrary hypercharge $y$ and $q$ is a rational number in the interval $(-1,1)$. This model has the global symmetry $U(1)_L^4 \times U(1)_R^4$, which corresponds to phase rotations of the $\psi$ and $\chi$ fields. Once the $SU(N_{\rm DC})$ gauge coupling becomes strong and the group confines, the global symmetry is spontaneously broken by the condensate $\langle \psi_i \chi_i\rangle \neq 0$ to $U(1)_V^4$. Thus there are four broken axial $U(1)'s$ which means there are also four GBs: one is eaten by the $U(1)_D$ gauge boson giving it a mass of $m_{\gamma_D} = 2(1-q) e_D f_{\rm DC}$ with $f_{\rm DC} \sim \Lambda_{\rm DC} /(4\pi)$, one is associated to an anomalous $U(1)_A$ under $SU(N_{\rm DC})$ (the $SU(N_{\rm DC})$ axion) and receives a large mass, one corresponds to a $U(1)_A$ anomalous under $SU(3)_c$ (the composite axion) and one is an exact GB. Note that the remaining GBs of the approximate $U(4)_L\times U(4)_R \rightarrow U(4)_V$, which is explicitly broken by the weak gauging of the SM gauge group, obtain a mass of the order $\tfrac{g_{\rm SM}^2}{(4\pi)^2} \Lambda_{\rm DC}^2$. The PQ symmetry current is of the form (see~\cite{Contino:2021ayn} for details)
\be
j^\mu_{\rm PQ} = \bar{\Psi} \gamma^\mu \gamma^5 Q_{\rm PQ} \Psi\,,\qquad Q_{\rm PQ}=\text{diag}(-3,-3,1,1)\,,
\ee
with $\Psi_L = (\psi_1, \psi_2,\psi_3,\psi_4)^T$ and $\Psi_R= (\chi_1^c, \chi_2^c, \chi_3^c,\chi_4^c)^T$. Note that $U(1)_{\rm PQ}$ does not have an $SU(N_{\rm DC})$ anomaly, i.e. $SU(N_{\rm DC})$ instantons do not contribute to the axion potential.

How can small QCD instantons contribute to the axion potential in this scenario? Below the confinement scale the composite axion behaves like an elementary axion and the discussion in Sections~\ref{sec:SmallInstanton} and~\ref{sec:QCDIR} fully applies. However, above the confinement scale of $SU(N_{\rm DC})$ the axion is not a relevant degree of freedom and $U(1)_{\rm PQ}$ is a linearly realized anomalous symmetry such that the QCD $\theta$-angle is unobservable (there are massless fermions in the spectrum). This implies that completely closing up all legs of the 't Hooft operator will not give a contribution to the axion potential above the confinement scale. 

However, instantons above the confinement scale can still generate effective operators, e.g. by closing only some of the legs of the 't Hooft operator. If this effective operator explicitly breaks the PQ symmetry and has a non-vanishing overlap with the axion after confinement it will give a contribution to the axion potential. This contribution is only misaligned with respect to the low-energy QCD contribution if a new CP violating phase enters the instanton computation. This can happen if we use CP violating higher-dimensional operators to close some of the legs of the 't Hooft operator.

To contribute to the axion potential the operator must have the same quantum numbers as the axion and in particular vanishing vectorial charges, i.e. the relevant PQ breaking operators can be written as polynomials of $(\psi_i \chi_i),(\psi_i \chi_i)^\dagger ,(\psi_i^\dagger \psi_i)$ or $(\chi_i^\dagger \chi_i)$~\cite{Contino:2021ayn}.

The 't Hooft operator itself already satisfies the above requirements: it explicitly breaks the PQ symmetry and it is a singlet under all vectorial $U(1)$'s. However, even after closing a few of its legs the result has to be gauge invariant and the gauge charge assignment of the new fermions greatly restricts the form of PQ breaking operators. 
\begin{figure}[t]
	\centering
	\includegraphics[width=0.5\textwidth]{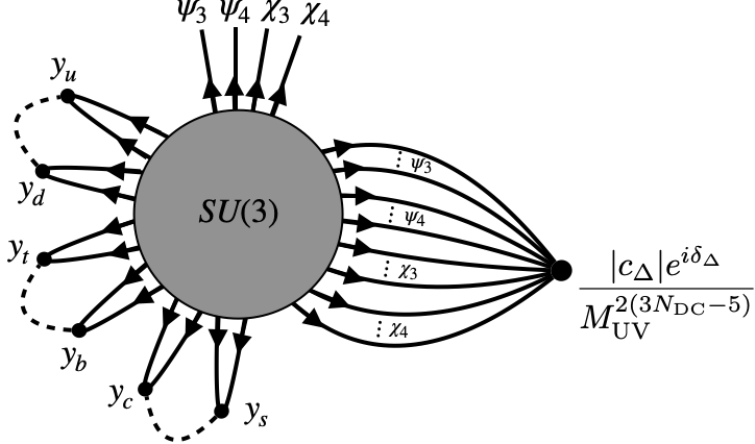}
	\caption{Contribution to the lowest-dimensional PQ breaking operator $\psi_3\psi_4 \chi_3\chi_4$ from QCD instantons above the confinement scale of $SU(N_{\rm DC})$, i.e. $\rho < 1/\Lambda_{\rm DC}$. The SM quark zero-modes are closed with Higgs loops, whereas the remaining $N_{\rm DC} - 1$ zero-modes for $\psi_3, \psi_4, \chi_3, \chi_4$ are assumed to be closed with one effective operator.}
	\label{fig:CompAxion}
\end{figure}
The lowest dimensional PQ breaking operators for the current example are $\psi_1\psi_2\chi_1\chi_2$ and $\psi_3\psi_4\chi_3\chi_4$. Instantons can only generate the second operator since all fermions in the first are color singlets. In order to close all legs of the 't Hooft operator in Figure~\ref{fig:CompAxion}, except for the ones in the PQ breaking operator $\psi_3\psi_4\chi_3\chi_4$, one can use Yukawa couplings and Higgs loops for the SM quarks and higher-dimensional operators for the remaining $N_{\rm DC}-1$ new fermions legs (see again Figure~\ref{fig:CompAxion}). Thus the contribution from $\rho \ll 1/\Lambda_{\rm DC}$ scales as 
\begin{align}
	\psi_3^{(0)}\psi_4^{(0)}\chi_3^{(0)}\chi_4^{(0)} |c_{\Delta}| e^{i \delta_{\Delta}} \frac{C_3}{(4\pi)^{\alpha}}  \left(\frac{8 \pi^2}{g^2}\right)^{6} \prod_{i=1}^6\frac{y_i}{4\pi}\int^{1/\Lambda_{DC}}_{1/M_{\rm UV}}\frac{d\rho}{\rho^5} \rho^6 \left(\Lambda_{SU(3)}\rho\right)^{b_0} \left( M_{\rm UV}\rho\right)^{-\Delta} 
\end{align}
where $\Delta$ counts the power of $1/M_{\rm UV}$ from the suppression scale of higher-dimensional operators and the $\psi_i^{(0)}$ and $\chi_i^{(0)}$ are fermion zero mode wavefunctions. If we use $n_{\mathcal{O}}$ operators to close the 't Hooft vertex this is given by $\Delta = 6(N_{\rm DC} -1 ) - 4 n_{\mathcal{O}}$ where we used that there are $4(N_{\rm DC} -1)$ zero-modes which have to be closed with higher-dimensional operators. The definition for $\alpha$ is given in Eq.~\eqref{eq:LoopFactor}. Here we restrict it to the zero-modes for the BSM fermions in which case it takes the form $\alpha = 4(N_{\rm DC} -1) - 2 n_{\mathcal{O}}$. In Figure~\ref{fig:CompAxion} we show an example where we use only one higher-dimensional operator to close $\psi_i$ and $\chi_i$ zero-modes. In this case $\Delta = 2(3N_{\rm DC} -5)$ and $\alpha = 2(2N_{\rm DC} - 3)$. The exact values depend on the number of operators we use to close the zero-modes but it will not affect our qualitative conclusions. $c_\Delta$ is the combination of Wilson coefficients used to close the zero-modes and $\delta_{\Delta}$ their combined phase. Due to the large number of colored particles also $b_0$ is modified to $b_0 = 7- 4N_{\rm DC}/3$. For $N_{\rm DC}\geq 3$ the integral is always UV dominated and one obtains
\begin{equation}
\begin{split}
    \frac{|c_{\Delta} | }{M_{\rm UV}^2} \frac{C_3}{(4\pi)^{\alpha}}  &\left(\frac{8 \pi^2}{g^2 (M_{\rm UV})}\right)^{6} \left(\frac{\Lambda_{SU(3)}}{M_{\rm UV}}\right)^{b_0}e^{i \delta_{\Delta}}\prod_{i=1}^6\left(\frac{y_i}{4\pi}\right) \psi_3^{(0)}\psi_4^{(0)}\chi_3^{(0)}\chi_4^{(0)} \\
    &= \frac{e^{-\tfrac{8\pi^2}{g^2(M_{\rm UV})}}}{M_{\rm UV}^2} \frac{|c_{\Delta} |}{(4\pi)^{2\Delta_{\rm UV}}} C_3 \left(\frac{8 \pi^2}{g^2 (M_{\rm UV})}\right)^{6} e^{i \delta_{\Delta}} \prod_{i=1}^6\left(\frac{y_i}{4\pi}\right) \psi_3^{(0)}\psi_4^{(0)}\chi_3^{(0)}\chi_4^{(0)}\,.
\end{split}
\end{equation}
For $g^2(M_{\rm UV}) < 8\pi^2$ this is always suppressed w.r.t. the natural expectation for the operator coefficient $\sim \mathcal{O}(1)/M_{\rm UV}^2$ from other possible UV dynamics. Note that in order to match this expression to an ordinary operator one has to use the explicit expression for the zero mode wavefunctions. However, this does not change the conclusion that the Wilson coefficient is exponentially suppressed for perturbative couplings. Thus effects from small instantons above the confinement scale seem to be negligible in composite axion models. However, below the confinement scale there can still be IR-dominated misaligned contributions to the axion potential along the lines of Section~\ref{sec:QCDIR}.

A simple way to summarize this, rather detailed, discussion is that UV instantons can pose a problem only if they generate the same PQ-breaking operators that the symmetries of~\cite{Contino:2021ayn} are designed to suppress or eliminate altogether. UV instantons respect these symmetries and so these models are structurally safe from the UV effects that we considered in this paper.
%
\section{Small Instantons in GUTs are Negligible (unless You Work Hard to Enhance Them)}
\label{sec:GUTs}
%

There are many models of GUTs which include an axion originating from a representation of the unified theory~\cite{Wise:1981ry,Reiss:1981nd,Mohapatra:1982tc,Holman:1982tb,Bajc:2005zf,Bertolini:2012im,Altarelli:2013aqa,Babu:2015bna,DiLuzio:2018gqe,Ernst:2018bib,Agrawal:2022lsp}. When studying instanton contributions to the axion potential in GUTs there are a few subtleties that one has to keep in mind (see also~\cite{Agrawal:2022lsp}):
\begin{itemize}
    \item The PQ symmetry is by definition anomalous under $SU(3)$. However, in GUTs the PQ symmetry is anomalous under the full unified gauge group which in particular contains both $SU(3)$ and $SU(2)_L$. This implies that the PQ symmetry must also be anomalous under $SU(2)_L$, i.e. the axion coupling at the GUT scale is
    \be
        \frac{a}{f_a} \frac{g_{\rm GUT}^2}{32\pi^2} \mathcal{G}^A_{\mu\nu}\tilde{\mathcal{G}}^{A\,\mu\nu} \supset \frac{a}{f_a} \left( r_3 \frac{g_{\rm GUT}^2}{32\pi^2} G^A_{\mu\nu}\tilde{G}^{A\,\mu\nu} + r_2 \frac{g_{\rm GUT}^2}{32\pi^2} W^a_{\mu\nu} \tilde{W}^{a\, \mu\nu}\right)\,,
    \ee
    where $r_3$ and $r_2$ denote the index of embedding of $SU(3)$ and $SU(2)$ into the GUT group, respectively. For a trivial embedding, as is the case for most simple GUTs, $r_3 = r_2 =1$. Thus both $SU(3)$ and $SU(2)_L$ instantons contribute to the axion potential. For UV dominated contributions their relative size scales with their instanton densities
    \be
        \frac{C_2 \left(\frac{2\pi}{\alpha_{\rm GUT}}\right)^4}{C_3 \left(\frac{2\pi}{\alpha_{\rm GUT}}\right)^6} \simeq 0.27\cdot \alpha_{\rm GUT}^2\,,
    \ee
    where we used the MSSM particle content in the instanton densities and that the couplings unify at the GUT scale. Thus while $SU(2)_L$ instantons contribute, their relative size is always subleading in perturbative GUTs. Note that here we imagine to ignore the UV theory and we perform our estimates in the SM EFT. However, this is one of the cases, mentioned at the end of Section~\ref{sec:NDA}, where Instanton NDA in the UV theory can give a larger effect than Instanton NDA in the EFT, if the result is UV-dominated. Nonetheless $SU(5)$ instantons do not change our  conclusion: you need to drastically change the SM $\beta$-functions before any of this becomes important.
    \item $SU(3)$ and $SU(2)_L$ unify into a simple group, so there is only one $\theta$ angle in the UV, i.e.
    \be
    \theta_{\rm QCD} = \theta_{\rm EW} = \theta_{\rm GUT}\quad \text{ at the GUT scale,}
    \ee
    for trivial embeddings. This means that $SU(2)_L$ instanton contributions to the axion potential are in general not misaligned with respect to QCD contributions.
\end{itemize}
Aside from these points that are  common to all GUTs, to compute the axion potential we have to make also a model-dependent choice. We have to specify an extension of the SM where gauge couplings unify.
As is well-known the SM gauge couplings do not unify exactly. Successful gauge coupling unification requires extra matter in SM representations between the TeV and GUT scale, possibly in the form of superpartners of the SM particles. In the MSSM, coupling unification works intriguingly well. 

Due to the requirement of extra matter charged under the SM gauge group it is natural to ask if small instantons can play an important role in generating the axion potential in such a setup.
In this section we approach this question from a bottom-up perspective and only add enough matter such that the couplings actually unify and assess how far this is from making small instantons important. For concreteness we focus on unification into $SU(5)$. 
%
\subsection{Split-SUSY Inspired GUT}\label{sec:splitSUSYGUT}
%
In order to achieve unification into $SU(5)$ Ref.~\cite{Giudice:2004tc} (see also~\cite{Giudice:2012zp}) considered the addition of vector like fermions to the SM which are part of the $\mathbf{5} + \mathbf{\bar{5}}$, $\mathbf{10} + \mathbf{\overline{10}}$, $\mathbf{15} + \mathbf{\overline{15}}$ or $\mathbf{24}$ representations of $SU(5)$. Under $(SU(3),SU(2)_L,U(1)_Y)$ these branch into
\begin{align}
Q&=(\mathbf{3},\mathbf{2},1 / 6)+(\mathbf{\overline{3}}, \mathbf{2},-1 / 6) & U&=(\mathbf{3},\mathbf{1},2 / 3)+(\mathbf{\overline{3}},\mathbf{1},-2 / 3) \\
D&=(\mathbf{3},\mathbf{1},-1 / 3)+(\mathbf{\overline{3}}, \mathbf{1} ,1 / 3) & L&=(\mathbf{1},\mathbf{2},1/2)+(\mathbf{1},\mathbf{2},-1/2) \\
E&=(\mathbf{1},\mathbf{1},1)+(\mathbf{1},\mathbf{1},-1) & V&=(\mathbf{1},\mathbf{3},0) \\
G&=(\mathbf{8},\mathbf{1},0) & X&=(\mathbf{3},\mathbf{2},-5 / 6)+(\mathbf{\overline{3}}, \mathbf{2},5 / 6) \\
T&=(\mathbf{1},\mathbf{3},1)+(\mathbf{1},\mathbf{3},-1) & S&=(\mathbf{6},\mathbf{1},-2 / 3)+(\mathbf{\overline{6}}, \mathbf{1} ,2 / 3)
\end{align}
One of the simplest choices which allows for unification is the addition of $(L+V+G)$. For TeV-scale vector like fermion masses this leads to unification at $M_{\rm GUT} = 1.6\cdot 10^{16}$~GeV with $\alpha_{\rm GUT}^{-1} = 35.9$ and predicts $\alpha_s (M_Z) = 0.102$ and therefore requires some threshold effects to correctly reproduce the PDG value of $\alpha_s (M_Z) = 0.1179(9)$~\cite{Workman:2022ynf}. $(L+V+G)$ is particularly interesting since it corresponds to the low-energy spectrum of split-supersymmetry~\cite{Arkani-Hamed:2004ymt}, i.e. 
light gauginos and higgsinos and heavy squarks and sleptons.\footnote{Note that the split-SUSY spectrum also contains a light bino, which is however not needed for coupling unification and therefore completely irrelevant for the further discussion.} The addition of $(L+V+G)$ modifies the $(SU(3),SU(2)_L,U(1)_Y)$ one-loop beta function coefficients to $b^{LVG} = \left(5,\tfrac{7}{6},-\tfrac{9}{2}\right)$. As we saw in Section~\ref{sec:misalignment} this is not sufficient for small instantons to give a sizable contribution to the axion potential. However, $b^{LVG}$ can easily be modified without spoiling coupling unification by adding additional matter in full $SU(5)$ representations. Let us first discuss how the additional fermions affect the instanton contribution to the axion potential in this setup before we mention possible modifications that would enhance small instantons.

We assume that $L+V+G$ have TeV scale vector like masses and make the connection to SUSY even more obvious by denoting the fermions by $L= \tilde{H}_d \oplus \tilde{H}_u$, $V = \tilde{W}$ and $G=\tilde{g}$. The most general renormalizable Lagrangian for the fermions is given by
\begin{equation}\label{eq:LVGrenLag}
	\begin{aligned}
-\mathcal{L}_{\rm ferm}= & y_{i j}^u \bar{q}_j u_i \epsilon H^*+y_{i j}^d \bar{q}_j d_i H+y_{i j}^e \bar{\ell}_j e_i H +\frac{M_G}{2} \tilde{g}^A \tilde{g}^A+\frac{M_V}{2} \tilde{W}^a \tilde{W}^a\\
&+\mu \tilde{H}_u^T \epsilon \tilde{H}_d +\frac{\tilde{g}_u}{\sqrt{2}}H^{\dagger} \sigma^a \tilde{W}^a \tilde{H}_u + h.c.-\frac{\tilde{g}_d}{\sqrt{2}}H^T \epsilon \sigma^a \tilde{W}^a \tilde{H}_d+h.c. \,,
\end{aligned}
\end{equation}
where $\mu \sim M_V\sim M_G \sim $~TeV. In split-SUSY $\tilde{g}_u$ and $\tilde{g}_d$ are related to the gauge couplings, but in our setup they are arbitrary $\mathcal{O}(1)$ Yukawa couplings for the new fermions. Note that terms of the form $y^{\ell\tilde{W}}_i  H^T \epsilon\, \sigma^a \tilde{W}^a \ell_i$ and $y^{\tilde{H}e}_i H^\dagger \tilde{H}_d \bar{e}_i$ do not appear in $R$-parity-conserving split-SUSY but they are consistent with all symmetries and originate from allowed $SU(5)$ Yukawa couplings. However, they violate lepton number and therefore have to be strongly suppressed. For this reason we will not consider them in the following.

For our purposes this renormalizable Lagrangian is not enough. As we have seen, instanton-generated 't Hooft operators can also be closed with effective operators. This is crucial for $SU(2)_L$ instantons which require an explicit breaking of $U(1)_{B+L}$ in order to give a contribution to the axion potential. $B+L$ violating operators, e.g. of the type $qqq\ell$, are  present in GUTs even without the inclusion of extra matter such as $L+V+G$. 

In order to estimate the size of these operators we need to have a class of UV-completions in mind. The most standard option consists in embedding the SM fermions $d, \ell$ in the $\mathbf{\bar{5}}$ and $q,u,e$ in the $\mathbf{10}$ representation of $SU(5)$. $V+G$ are part of the $\mathbf{24}$ and $L$ of the vector like pair $\mathbf{5}+\mathbf{\bar{5}}$ since integrating them out leads to higher-dimensional operators with two fermions and two scalars which require additional scalar loops to close the 't~Hooft operator. Assuming for now that there is only one scalar field $H_5 = (T,H)$ in the $\mathbf{5}$ representation, containing the light SM Higgs $H$ and a heavy colored Higgs $T$, the following Yukawa interactions exist in the $SU(5)$ symmetric theory  
\begin{align}
	(y_u)^{ij} 10_i 10_j H_5 &\supset (y_u)^{ij} \left[q_i \epsilon H \bar{u}_j + T\bar{u}_i \bar{e}_j + q_i \epsilon q_j T\right]\,,\\
	(y_d)^{ij} \bar{5}_{i} 10_j(H_5^\ast)_\beta &\supset (y_d)^{ij}\left[ \bar{d}_i \bar{u}_j T^\ast + \ell_i q_j T^\ast + H^\dagger q_i \bar{d}_j + H^\dagger \ell_i \bar{e}_j\right]\,,\\
 (y_1)^i \bar{5}_L 10_i H_5^* &\supset (y_1)^i\left[q_i\epsilon \tilde{H}_d T^\ast + H^\dagger \tilde{H}_d \bar{e}_i\right]\,,\\
	y_2 \bar{5}_L 24_{VG} H_5 &\supset y_2 \tilde{H}_d \tilde{W}^a \sigma^a H\,, \\
	y_3 H_5^\dagger 24_{VG}\, 5_L &\supset y_3 H^\dagger \tilde{W}^a \sigma^a \tilde{H}_u\,, 
\end{align}
where gauge index contractions are left implicit and fermion fields are denoted by the dimension of their representation under $SU(5)$. The heavy vector like fermions $L+V+G$ are embedded into the $SU(5)$ multiplets $5_L+ \bar{5}_L$ and $24_{VG}$. Note that we have not added a Yukwawa coupling of the form $(y_4^i )\bar{5}_i\, 24 H_5$ which would produce the lepton number violating operators mentioned below Eq.~\eqref{eq:LVGrenLag}.
The couplings $(y_u)^{ij}$ and $(y_d)^{ij}$ are the SM Yukawa matrices. We have left out couplings including heavy fermions in the $\mathbf{24}$ and $\mathbf{5}_L,\mathbf{\bar{5}}_L$. Integrating out the colored triplet Higgs $T$ one arrives at dimension six operators of the form
\begin{equation}\label{eq:GUTeffOp}
\begin{split}
	\mathcal{L}_6 \supset &\frac{c_{qqq\ell}^{ijkm}}{m_T^2} q_i q_j q_k \ell_m  + \frac{c_{qqdu}^{ijkm}}{m_T^2} q_i q_j \bar{u}_k \bar{d}_m  + \frac{c_{qqq\tilde{H}}^{ijk}}{m_T^2} q_i \epsilon q_j q_k \epsilon \tilde{H}_d\,,  
	\end{split}
\end{equation}
where we only kept operators where all fermions are charged under either $SU(3)$ or $SU(2)_L$. The Wilson coefficients for these operators scale as $c_{qqq\ell},c_{qqdu} \propto y_u y_d$ and $c_{qqq\tilde{H}} \propto y_1 y_u$
with $y_1,y_2,y_3 \sim \mathcal{O}(1)$ in general. Note that $y_2,y_3$ can be identified with $\tilde{g}_d,\tilde{g}_u$ in Eq.~\eqref{eq:LVGrenLag}.

\begin{figure}[t]
	\centering
	\includegraphics[width=\textwidth]{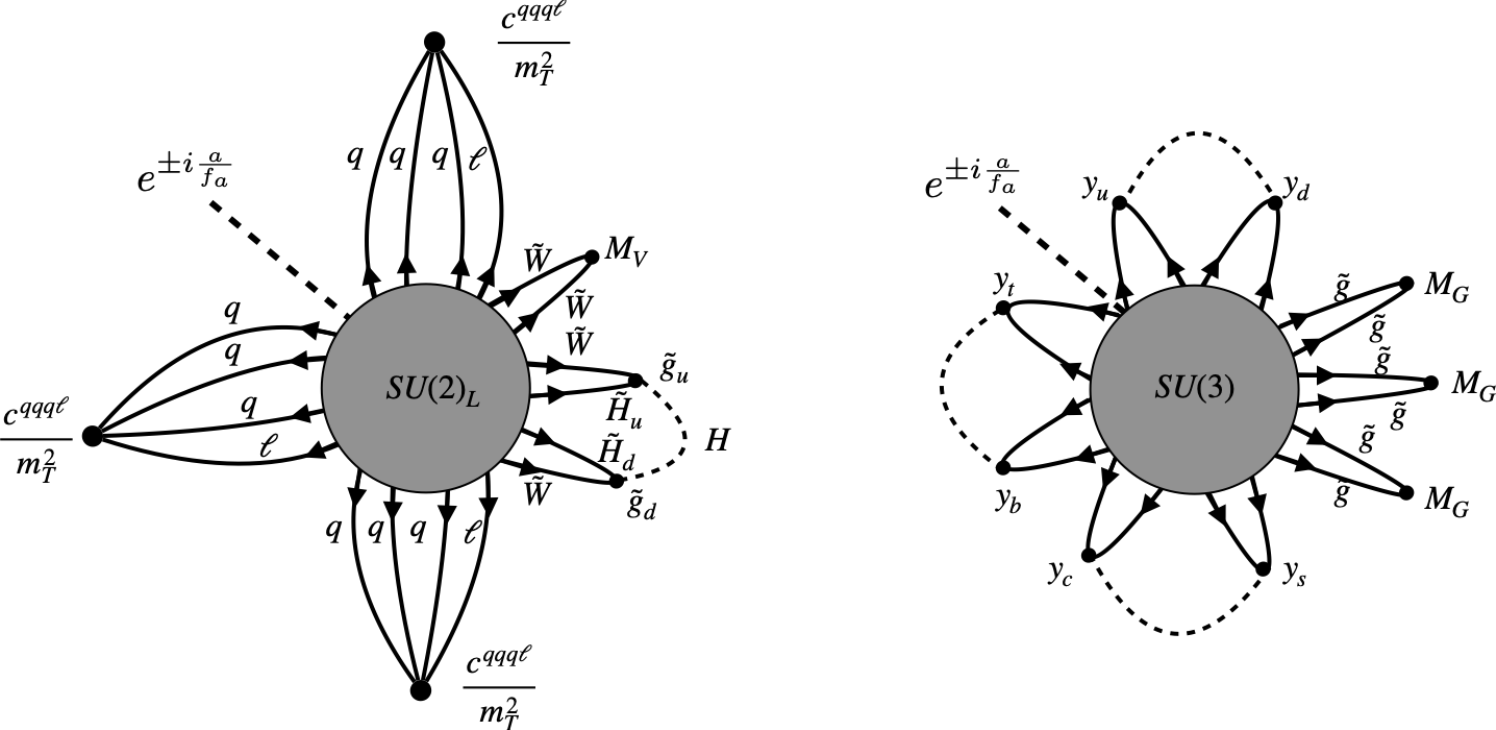}
	\caption{Contributions to the axion potential from $SU(2)_L$ and $SU(3)$ instantons using the couplings in Eq.~\eqref{eq:LVGrenLag} and Eq.~\eqref{eq:GUTeffOp}.}
	\label{fig:GUTEFT}
\end{figure}
Using the couplings in Eq.~\eqref{eq:LVGrenLag} and Eq.~\eqref{eq:GUTeffOp} we can close the $SU(2)_L$ and $SU(3)$ 't Hooft operators as shown in Figure~\ref{fig:GUTEFT} and obtain a contribution to the axion potential. Note that in addition to the zero-modes for the SM fermions there are also zero-modes for $L+V+G$. $V$ and $G$ transform in the adjoint representations of $SU(2)_L$ and $SU(3)$ and have four and six zero-modes, respectively. In order to close the zero-modes of the fermions in the adjoint representation mass insertions are required. Using the power counting rules of Section~\ref{sec:InstantonCalculus}, we find that the contributions from $SU(2)_L$ instantons, depicted in the left panel of Figure~\ref{fig:GUTEFT}, scale as
\be \label{eq:GUTSU2}
\Lambda_{{\rm SI},\, SU(2)_L}^4 \simeq 2 C_2 \left(\frac{8\pi^2}{g_w^2(M_{\rm GUT})}\right)^4 \left(\frac{c_{qqq\ell}}{16\pi^2}\right)^3 \frac{\tilde{g}_u \tilde{g}_d}{(4\pi)^2} \left( \frac{\Lambda_{SU(2)}^{(LVG)}}{M_{\rm GUT}}\right)^{b_{SU(2)}^{LVG}-4} \left(\frac{M_V}{M_{\rm GUT}}\right) \Lambda_{SU(2)}^{(LVG)}\,^4\,.
\ee
The integral over the instanton size is UV dominated and we have taken $m_T \simeq M_{\rm GUT}$. The RGE invariant scale is given by $(\Lambda_{SU(2)}^{(LVG)})^{b_{SU(2)}^{LVG}} = M_{\rm GUT}^{b_{SU(2)}^{LVG}} e^{-\frac{2\pi}{ \alpha_{\rm GUT}}}$ and evaluates to $\Lambda_{SU(2)}^{(LVG)} \simeq 1.7\cdot 10^{-68}$~GeV. Taking $\tilde{g}_u \sim \tilde{g}_d \sim \mathcal{O}(1)$ and $c_{qqq\ell} \sim y_u y_d$ this leads to $\Lambda_{{\rm SI},\, SU(2)_L} \simeq 1.7\cdot 10^{-16}\text{ GeV}\ll \Lambda_{\rm QCD}$.

The perturbative QCD contribution, from instantons of size $\rho < M_G^{-1}$, depicted in the right panel of Figure~\ref{fig:GUTEFT} is dominated by its lowest scale $\rho\simeq M_G^{-1}$, and gives a contribution that is largely subdominant to IR QCD dynamics at scales $\mathcal{O}(\Lambda_{\rm QCD})$. This can be seen by applying our power-counting rules to the right panel of Figure~\ref{fig:GUTEFT},
\be \label{eq:SU3GUT}
\Lambda_{{\rm SI},\, SU(3)}^4 \simeq 2 C_3 \left(\frac{8\pi^2}{g^2(M_G)}\right)^6 \prod_i \frac{y_i}{(4\pi)}\left( \frac{\Lambda_{SU(3)}^{(LVG)}}{M_{G}}\right)^{b_{SU(3)}^{LVG}-4}  \Lambda_{SU(3)}^{(LVG)}\,^4\,,
\ee
where we used the mass threshold $M_G$ as the IR cutoff. With the RGE invariant scale $\Lambda_{SU(3)}^{(LVG)} \simeq 4.1\cdot 10^{-4}$~GeV this evaluates to $\Lambda_{{\rm SI},\, SU(3)} \simeq 4.4\cdot 10^{-9}\text{ GeV} \ll \Lambda_{\rm QCD}$.

In this setup the suppression of small instantons does not only originate from a small gauge coupling at the GUT scale, but also from the TeV-scale mass insertions required to close zero-modes of fermions in the adjoint representation. Note that this is also a feature of supersymmetric theories where at least one insertion of the supersymmetry breaking scale, corresponding to the masses of the adjoint fermions in our model, is needed to generate a potential for the axion~\cite{Dine:1986bg,Dine:2022mjw}. 

In the present case the number of mass insertions for QCD instanton contributions can easily be reduced by introducing a real singlet scalar $\phi$ with a Yukawa coupling to the $\mathbf{24}$ representation that $\tilde{W}$ and $\tilde{g}$ are embedded in, i.e.
\be \label{eq:yukawa24}
y_{24}\, \phi\, 24\, 24 \supset y_{24}\, \phi\, \left( \tilde{W}^a \tilde{W}^a + \tilde{g}^A \tilde{g}^A \right)\,.
\ee
If $\phi$ is light the zero-modes can be closed with a $\phi$ loop, removing the extra $M_G/M_{\rm GUT}$ suppression. If it is heavy, for instance because of a (more natural) GUT scale mass, it can be integrated out producing four-fermions contact interactions that can close the zero-modes, giving parametrically the same result as of a light $\phi$.

However, note that one TeV-scale mass insertion is unavoidable since both for $SU(2)_L$ and $SU(3)$ there are six zero-modes of TeV-scale fermions and only four of them can be closed with Yukawa couplings or an effective four-fermion operator.

\begin{figure}[t]
	\centering
	\subfigure{\includegraphics[width=0.298\textwidth]{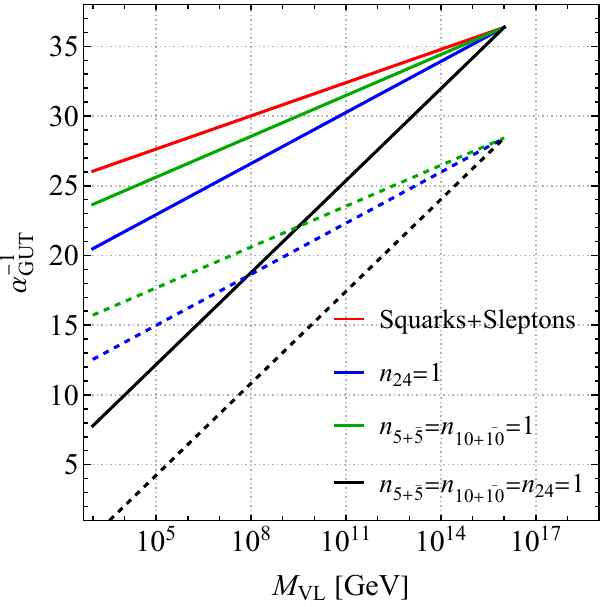}}\hfill
	\subfigure{\includegraphics[width=0.345\textwidth]{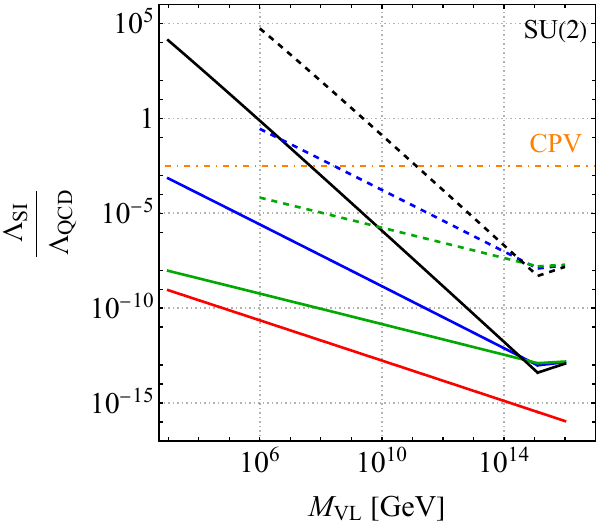}}
    \subfigure{\includegraphics[width=0.34\textwidth]{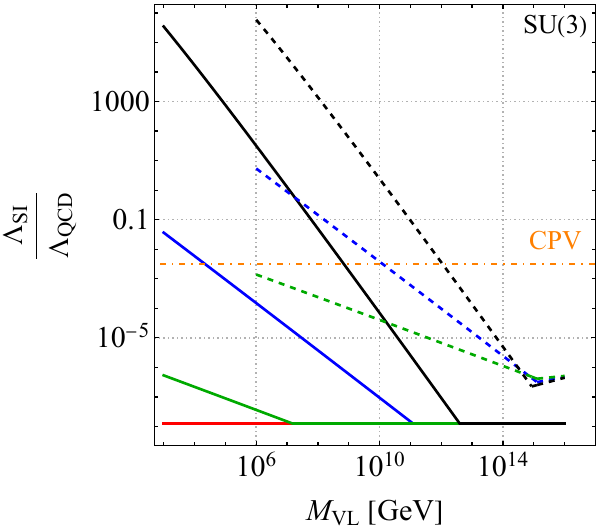}}
	\caption{GUT coupling (left) and small instanton contribution to the axion potential relative to the low-energy QCD contribution for $SU(2)$ (middle) and $SU(3)$ (right) as a function of the scale $M_{\rm VL}$ at which new vector like fermions or superpartners appear. Dashed lines assume the mini split-SUSY spectrum with sleptons and squarks with masses of the order $100$~TeV in addition to the vector like fermions at the mass scale $M_{\rm VL} > 100$~TeV. The orange dot-dashed line shows the maximally allowed contribution that does not spoil the solution to the strong-CP problem if a new source of CP violation enters the small instanton contribution.}
	\label{fig:GUTModification}
\end{figure}

Let us now discuss a few simple modifications of the model which might change the above conclusions and make UV instanton contributions sizable. The size of the contribution to the axion potential in all these setups compared to the low-energy QCD contribution is shown in Figure~\ref{fig:GUTModification}.
\begin{itemize}
	\item {\bf Scalar Superpartners:} A well motivated addition to the particle spectrum are the scalar superpartners of the SM fermions, i.e. the squarks and sleptons. In mini-split SUSY (see e.g.~\cite{Arvanitaki:2012ps}) the scalars have masses which are about a loop factor larger than the fermion masses. The squarks and sleptons change the $SU(3)$ and $SU(2)_L$ beta functions by $\Delta b_3 = -2$ and $\Delta b_2 = -13/6$, respectively. 
 The addition of the superpartners alone is however not enough to get a sufficiently large coupling at the GUT scale (see Figure~\ref{fig:GUTModification} left), but in combination with further $SU(5)$ symmetric matter a sizable coupling can be reached.
	\item {\bf 24 + real scalar $\phi$:} The arguably simplest option is to add an additional set of fermions in the adjoint representation of $SU(5)$ $\widetilde{24}$ with a real scalar $\phi$. This allows us to add the following term to the Lagrangian 
		\begin{equation}
			\left( \frac{\widetilde{m}_{24}}{2} + \tilde{y}_{24} \phi \right) \widetilde{24}^A \widetilde{24}^A\,.
		\end{equation}
		Adding a $24$-plet implies that there are $10$ new zero modes both for $SU(3)$ and $SU(2)_L$.\footnote{Under $(SU(3),SU(2)_L)_{U(1)_Y}$ the $\mathbf{24}$ representation of $SU(5)$ branches into $\mathbf{24}= (\mathbf{8},\mathbf{1})_0 + (\mathbf{1},\mathbf{3})_0 + (\mathbf{1},\mathbf{1})_0 + (\mathbf{3},\mathbf{2})_{-5/6}+ (\mathbf{\bar{3}},\mathbf{2})_{5/6}$.} The Yukawa coupling can be used to close eight of them with $\phi$ loops, so that one only has to pay the loop factor suppression. If $\phi$ also couples to $\tilde{g}^A$ and $\tilde{W}^a$ as in Eq.~\eqref{eq:yukawa24} all zero-modes from adjoint fermions can be closed. The $24$-plet contributes with $\Delta b_3 = \Delta b_2 = - 10/3$ to the beta functions. The size of the corresponding contribution to the axion potential as a function of the vector like fermion mass is shown in blue in Figure~\ref{fig:GUTModification}.
	\item {\bf Vector like complex representations + real scalar $\phi$:} Another possibility is to add an even number of vector like fermions in a complex representations such as e.g. $\mathbf{5}+\mathbf{\bar{5}}$ and $\mathbf{10}+\mathbf{\overline{10}}$ with a real scalar $\phi$ that couples to the fermions as
		\begin{equation}
			y_5 \phi\, \bar{5}\, 5 + y_{10} \phi \overline{10}\, 10\,. 
		\end{equation}
		 This adds $2n_{5+\bar{5}} + 6n_{10+\overline{10}}$ new zero modes for $SU(2)_L$ and $SU(3)$.\footnote{Under the SM gauge group these branch into $\mathbf{5}+\mathbf{\bar{5}}= (\mathbf{\bar{3}},\mathbf{1})_{1/3} + (\mathbf{3},\mathbf{1})_{-1/3} + (\mathbf{1},\mathbf{2})_{1/2}+ (\mathbf{1},\mathbf{2})_{-1/2}$ and $\mathbf{10}+\mathbf{\overline{10}} = (\mathbf{\bar{3}},\mathbf{1})_{-2/3} + (\mathbf{3},\mathbf{1})_{2/3} + (\mathbf{1},\mathbf{1})_{-1}+ (\mathbf{1},\mathbf{1})_{1} + (\mathbf{3},\mathbf{2})_{1/6}+ (\mathbf{\bar{3}},\mathbf{2})_{-1/6}$.} If $n_{5+\bar{5}} + n_{10+\overline{10}}$ is even they can be closed with $\phi$ loops or effective operators that one would obtain when integrating out $\phi$ in case it is heavy. Figure~\ref{fig:GUTModification} shows $n_{5+\bar{5}} = n_{10+\overline{10}} = 1$ in green.
		
	    Instead of introducing the real scalar $\phi$ one can also add a full vector like fourth family, in which case the SM Yukawa couplings can be replicated for this new family, such that the zero modes can be closed with Higgs loops in the $SU(3)$ case or the effective operators obtained by integrating out the color triplet Higgs $T$ in the $SU(2)_L$ case. This changes the beta functions by $\Delta b_i = -2\cdot n_{10+\overline{10}} - \tfrac{2}{3} n_{5+\bar{5}}$.
\end{itemize}
The GUT coupling and the instanton contribution to the axion potential for the above cases is shown in Figure~\ref{fig:GUTModification} as a function of the scale where the new particles appear. Note that the $SU(3)$ instanton contribution in Figure~\ref{fig:GUTModification} has a lower bound. This originates from the IR dominated contribution between $1/M_{\rm VL} < \rho < 1/M_G$ given in Eq.~\eqref{eq:SU3GUT} which is always present. The kink both in the $SU(2)_L$ and $SU(3)$ contributions occours at $M_{\rm GUT} / (4\pi)$ where it becomes favorable to close the zero-modes of the new vector like fermions with a $M_{\rm VL}$ mass insertion instead of Yukawa couplings and a $\phi$ loop. For the parametric scaling of the instanton contributions see Appendix~\ref{app:GUTInstanton}.
In conclusion the supersymmetric particle spectrum alone is not enough to make small instantons important. However, as in all previous sections this is a UV dependent statement as additional matter in $SU(5)$ representations can modify the running of the gauge coupling. Depending on the CP structure of the GUT this would either enhance the axion mass (no additional CP vioalation) or spoil the axion solution to the strong CP problem (new CP violating phases enter the instanton calculation).
%
\section{Conclusions} \label{sec:conclusions}
In this work we presented simple power counting rules that allow us to instantly evaluate vacuum-to-vaccum amplitudes in an instanton background up to $\mathcal{O}(1)$ factors. The six steps, listed in Section~\ref{sec:NDA}, are as simple as what one would normally use to estimate the contribution of a Feynman diagram to the amplitude of a physical process. We called these rules {\it Instanton NDA}, as their application is rather similar to that of regular NDA. The basic conceptual steps of an instanton calculation and the  process that leads to this set of rules are explained in Section~\ref{sec:InstantonCalculus}.

These rules constitute a considerable simplification compared to the standard procedure of evaluating instanton effects. 
Ordinarily, computing a vacuum-to-vacuum amplitude in an instanton background requires: 1) Computing the zero-mode wavefunctions of all fields charged under the gauge group and the solutions of the equations of motion in an instanton background, 2) Expanding the path integral around these solutions at least up to quadratic order in the fluctuations, 3) Performing the  Gaussian path integral over the non-zero modes of the fields 4) Integrating the zero-mode wavefunctions over the instanton size and orientation within the gauge group. This procedure is obviously cumbersome and unsuitable for a quick order-of-magnitude estimate. Additionally, it is often impossible to complete it analytically, either because the zero-mode wavefunctions or propagators are not known for arbitrary representations or because the integral over instanton sizes can only be done numerically. The simple rules presented here bypass these difficulties and streamline the process of estimating  instanton effects.

As interest in axion physics is steadily growing, performing such estimates is becoming increasingly relevant. The axion potential in QCD is dominated by confining dynamics in the IR, as reviewed in Section~\ref{sec:lowE}, and an instanton calculation cutoff at $\Lambda_{\rm QCD}$ would give the wrong result. However, instantons encode correctly high-energy contributions when the gauge coupling is perturbative. These contributions  can be important both for the size of the potential and that of the neutron EDM. 

In Sections~\ref{sec:examples}, \ref{sec:misalignment}, \ref{sec:UVSafe}, and \ref{sec:GUTs} we applied Instanton NDA to the calculation of UV effects on the axion potential. In Section~\ref{sec:examples} we discussed, as a warm up exercise, three toy examples that capture the three physically distinct ways of generating an axion potential from instantons. In Section~\ref{sec:misalignment} we emphasized the dangers posed by enhancing the UV contributions to the axion mass. Enhancing these effects also enhances the neutron EDM and make it sensitive to high-energy sources of CP violation. The peculiarity of instanton effects is that they do not always decouple as the UV scale of CP violation goes to infinity. This is an important qualitative feature that emerges immediately from Instanton NDA and allows to asses whether a model is UV-safe or not. Two UV-safe models are discussed in detail in Section~\ref{sec:UVSafe}. We conclude our list of applications of Instanton NDA in Section~\ref{sec:GUTs}, by showing that UV effects in GUTs are highly subdominant to the QCD potential in the IR. We discuss a number of possible ways of enhancing UV effects, but they all require the addition of many new matter multiplets.

In conclusion, we have introduced a simple but powerful tool to streamline estimates of instanton calculations that greatly reduces the burden of evaluating UV contributions to the axion potential. We then discussed in detail multiple possible applications, ranging from GUTs to misalignment from high-energy CP violation.  

%
\section{Acknowledgements} 
 We thank Pablo Sesma for useful comments and discussions. CC and MR are supported in part by the NSF grant PHY-2014071. MR is also supported by a Feodor–Lynen Research Fellowship awarded by the Humboldt Foundation.   CC and EK are funded in part by the US-Israeli BSF grant 2016153. CC also thanks the Aspen Center for Physics (supported by the NSF grant PHY-2210452) where part of this work was done. 

%
\appendix
%
\section{Loop Factor Counting for Instanton NDA} \label{app:LoopFactor}
%
In this appendix we justify the loop factor counting for closed 't Hooft operators as given in Eq.~\eqref{eq:LoopFactor}. The contribution to the vacuum energy is computed according to Eq.~\eqref{eq:SUNInstanton}. This entails projecting out fermion zero modes from mass or interaction terms and a path integral over all remaining dynamical fields. As can be seen in the example in Eq.~\eqref{eq:InstantonYukawa} this reduces to a perturbative expansion in the couplings up to an order that is sufficient to provide one fermion field for each zero mode. All additional fields that appear in the interactions only survive the path integral if they can be fully Wick-contracted, yielding propagators. Up to $\mathcal{O}(1)$ factors these yield the following factors of $\pi$
\be
\text{zero-mode $\psi^{(0)}$} \sim \frac{1}{\pi}\,,\qquad \text{vertex } \sim \pi^2\,,\qquad \text{propagator $\Delta_F$}\sim \frac{1}{\pi^2}\,,
\ee
where we used that the explicit expression for the zero mode in Eq.~\eqref{eq:FermZeroMode} is $\psi^{(0)}\propto 1/\pi$, each vertex comes with an integral over spacetime $\int d^4x \propto \pi^2 \int dx\, x^3$ and the propagator $\Delta_F$ contains an integral over all momenta $\int\frac{d^4p}{(2\pi)^4} \propto \frac{1}{\pi^2}\int dp\, p^3$. Writing this in powers of the conventional NDA loop factor $(4\pi)$ one arrives at Eq.~\eqref{eq:LoopFactor}.
%
\section{Enhanced Axion Mass from Instantons in Partially Broken Gauge Groups} \label{app:InstIOE}
%
In this appendix we demonstrate how embedding QCD into a larger gauge group $G'$ in the UV with a non-trivial index of embedding enhances UV instanton contributions to the axion potential. In such a case instantons in the UV gauge group are not matched one-to-one to low-energy instantons (see e.g.~\cite{Intriligator:1994jr,Intriligator:1994sm,Intriligator:1995id,Csaki:1998vv,Csaki:2019vte}). An index of embedding $r$ means that a one-instanton effect in the IR theory corresponds to a $r$-instanton effect in the UV theory. This implies that one-instanton effects in the UV theory effectively scale like $1/r$ fractional instantons and have to be integrated out when matching to the IR theory. The instanton matching is taken into account through a modified scale-matching relation of the form~\cite{Csaki:1998vv}
\be \label{eq:ScaleMatchingNonTrivial}
\left(\frac{\Lambda_{G}^{\rm IR}}{M}\right)^{b_0^{\rm IR}} = \left(\frac{\Lambda_{G'}}{M}\right)^{r\cdot b_0}\,,
\ee
where $r$ is the index of embedding and $M$ the matching scale, where we now assume $M \gg \Lambda_G^{\rm IR}$. Let us demonstrate this effect using one of the simplest cases where a non-trivial index of embedding occurs. If we assume that the $SU(N)$ gauge group of the examples we considered in Section~\ref{sec:EnhancedSmallInstantons} is the diagonal combination of a product gauge group $SU(N)^r$ which is spontaneously broken by the VEV $v$ of a scalar, one-instanton configurations in any of the $SU(N)$ factors in the UV have no analog in the IR. A one-instanton configuration in the IR gauge group is a simultaneous $(1,1,\ldots, 1)$ instanton in the $r$ factors, i.e. the index of embedding is $r$ (see e.g.~\cite{Agrawal:2017ksf,Agrawal:2017evu,Csaki:2019vte}). In order to be more concrete we assume that the two vector like fermions in the previous result couple to the first factor of the product gauge group. In this case the one-instanton configuration within the first factor sits completely in the broken part of the gauge group and has a natural IR cutoff due to the exponential suppression of instantons which are much larger than $1/v$. Thus if we can close the zero-modes with loops of scalars, the instanton effect scales as
\be \label{eq:InstantonHiggsEmbedding}
\begin{split}
\vcenter{\hbox{\includegraphics[width=0.25\textwidth]{Figures/HiggsPotential.pdf}}} &\sim C_N \left(\frac{8\pi^2}{g^2}\right)^{2N} e^{\pm i \frac{a}{f_a}}\int_{1/M_{\rm UV}}^{\infty}\frac{d\rho}{\rho^5} (\Lambda_{SU(N)} \rho)^{b_0} e^{-2\pi^2 \rho^2 v^2} \frac{y_1 y_2}{16\pi^2}\\
&\sim C_N \left(\frac{8\pi^2}{g^2(2\pi v)}\right)^{2N} e^{\pm i \frac{a}{f_a}} \frac{y_1 y_2}{16\pi^2} \left(\frac{\Lambda_{SU(N)}}{2\pi v}\right)^{b_0} \left(2\pi v\right)^4\\
&\sim C_N \left(\frac{8\pi^2}{g^2 (2\pi v)}\right)^{2N} e^{\pm i \frac{a}{f_a}} \frac{y_1 y_2}{16\pi^2} \left(\frac{g (2\pi v)}{2\pi}\right)^{b_0 - 4} \left(\frac{\Lambda_{SU(N)}^{\rm IR}}{M}\right)^{b_0/r -4} (\Lambda_{SU(N)}^{\rm IR})^4
\end{split}
\ee
where we dropped $\mathcal{O}(1)$ factors and took into account that the physical matching scale associated to particle masses is of the order $M\sim g v$. As can be seen, already for an index of embedding of $r=2$ there is an enhancement for $b_0 <8$.\footnote{Note that for product gauge groups the matching relation is $\left(\frac{\Lambda^{\rm IR}_G}{M}\right)^{b_0^{\rm IR}} = \prod_{i=1}^r \left(\frac{\Lambda_G^{(i)}}{M}\right)^{b_0^{(i)}}$ which simplifies to Eq.~\eqref{eq:ScaleMatchingNonTrivial} if all group factors are identical, which is what we assumed for simplicity.} For $r\gg 1$ one achieves the maximal enhancement and an axion potential that scales as $V_a \sim M^4$, with no trace of the IR contribution remaining.
%
\section{Instanton Contributions to the Axion Potential in GUTs} \label{app:GUTInstanton}
%
Here we collect the expressions for the instanton contributions to the axion potential for the GUT modifications discussed in Section~\ref{sec:splitSUSYGUT} and plotted in Figure~\ref{fig:GUTModification}. In all of these expressions we assume that a Yukawa coupling of the form Eq.~\eqref{eq:yukawa24} for the adjoint fermions exists, which can be beneficial to avoid mass suppressions. Note that here we only consider contributions from instantons of sizes $4\pi/M_{\rm GUT} < \rho < 1/M_{\rm VL}$. For $SU(3)$ there still exists an IR dominated contribution from $1/M_{\rm VL} < \rho < 1/M_G$, which will dominate if the contribution from $4\pi/M_{\rm GUT} < \rho < 1/M_{\rm VL}$ is smaller. This can be seen in the right panel of Figure~\ref{fig:GUTModification}. For $1/M_{\rm GUT} < \rho < 4\pi / M_{\rm GUT}$ it is favorable to close zero modes of the new vector like fermions with $M_{\rm VL}$ mass insertions instead of Yukawa couplings and $\phi$ loops.
\paragraph{Scalar superpartners}~\\
The addition of scalar superpartners is not enough to make the $SU(3)$ instanton UV dominated. The contribution from $M_G < 1/\rho < M_{\rm VL}$ given in Eq.~\eqref{eq:SU3GUT} is still the leading one from small instantons. Using the Yukawa couplings from Eq.~\eqref{eq:yukawa24} to close the zero modes of the adjoint fermion is subleading for $\mathcal{O}(1)$ Yukawa couplings due to the IR dominance of the instanton size integral and the loop suppression.

For $SU(2)_L$ instantons the Yukawa couplings cannot prevent one TeV-scale mass insertion. The dominant contribution is therefore still given by Eq.~\eqref{eq:GUTSU2} with the replacement $b_{SU(2)}^{LVG} \rightarrow b_{SU(2)}^{SUSY}$ and $\Lambda_{SU(2)}^{(LVG)} \rightarrow \Lambda_{SU(2)}^{SUSY}$ with $b_{SU(2)}^{SUSY}=-1$. Thus there is an enhancement from the larger gauge coupling at the GUT scale.
\paragraph{24 + real scalar}~\\
The addition of a full $\mathbf{24}$-plet changes the beta functions by $\Delta b_i = -10/3$ which is enough to make the contribution from $SU(3)$ instantons UV dominated for $\rho < 1/M_{\rm VL}$ if we close the zero-modes with Yukawa couplings. This can be done for the zero-modes of all new particles and a pair of $\tilde{g}^A$ with $\phi$ loops. This results in an expression of the form
\be
\Lambda_{\rm SI}^4 \simeq 2 C_3 \left( \frac{8\pi^2}{g^2(M_{\rm GUT})}\right)^6 \left(\prod_i \frac{y_i}{4\pi}\right) \frac{y_{24}^3 \tilde{y}_{24}^5}{(4\pi)^{8}}\left(\frac{\Lambda_{SU(3)}}{M_{\rm GUT}}\right)^{b^{SU(3)}-4} \Lambda_{SU(3)}^4\,,
\ee
with $b^{SU(3)} = 5/3$ and $\Lambda_{SU(3)} = M_{\rm GUT} \exp\left[-2\pi /(\alpha_{\rm GUT}\, b^{SU(3)}) \right]$, where $\alpha_{\rm GUT}$ depends on $M_{\rm VL}$ and is shown in the left panel of Figure~\ref{fig:GUTModification}.

Contributions from $SU(2)_L$ instantons are UV dominated and analogously to the $SU(3)$ case above all new zero-modes and one pair of $\tilde{W}^a$ zero-modes can be closed with $\phi$ loops. The corresponding scale generated by instantons is given by
\be
\Lambda_{\rm SI}^4 \simeq 2 C_2 \left( \frac{8\pi^2}{g_w^2(M_{\rm GUT})}\right)^4 \left(\frac{c_{qqq\ell}}{16\pi^2}\right)^3 \frac{\tilde{g}_u \tilde{g}_d}{(4\pi)^2} \frac{y_{24} \tilde{y}_{24}^5}{(4\pi)^6} \left(\frac{\Lambda_{SU(2)}}{M_{\rm GUT}}\right)^{b^{SU(2)}-4} \Lambda_{SU(2)}^4\,,
\ee
with $b^{SU(2)} = -13/6$ and $\Lambda_{SU(2)}$ defined in the same way as for $SU(3)$.
\paragraph{Vector like complex representations + real scalar $\phi$}~\\
If we add $n_{5+\bar{5}}=n_{10+\overline{10}}=1$ fermions to the theory the beta functions get modified by $\Delta b_i = -8/3$. Since we added eight zero-modes both to $SU(3)$ and $SU(2)_L$ we can no longer close all zero-modes with Yukawa couplings and $\phi$ loops but need at least one mass insertion. It is favorable to use the mass insertion for one of the vector like fermions and not for $\tilde{g}^A$ or $\tilde{W}^a$. For $SU(3)$ this yields
\be
\Lambda_{\rm SI}^4 \simeq 2 C_3 \left( \frac{8\pi^2}{g^2(M_{\rm GUT})}\right)^6 \left(\prod_i \frac{y_i}{4\pi}\right) \frac{y_{24}^3 y_{10}^3}{(4\pi)^{6}}\left(\frac{M_{\rm VL}}{M_{\rm GUT}} \right)\left(\frac{\Lambda_{SU(3)}}{M_{\rm GUT}}\right)^{b^{SU(3)}-4} \Lambda_{SU(3)}^4\,,
\ee
with $b^{SU(3)} = 7/3$. For $SU(2)_L$ one finds
\be
\Lambda_{\rm SI}^4 \simeq 2 C_2 \left( \frac{8\pi^2}{g_w^2(M_{\rm GUT})}\right)^4 \left(\frac{c_{qqq\ell}}{16\pi^2}\right)^3 \frac{\tilde{g}_u \tilde{g}_d}{(4\pi)^2} \frac{y_{24} y_{10}^3}{(4\pi)^4} \left(\frac{M_{\rm VL}}{M_{\rm GUT}} \right) \left(\frac{\Lambda_{SU(2)}}{M_{\rm GUT}}\right)^{b^{SU(2)}-4} \Lambda_{SU(2)}^4\,,
\ee
where $b^{SU(2)} = -3/2$. In both cases we chose to close the zero-modes of $\mathbf{5}+\mathbf{\bar{5}}$ with a mass insertion.
\paragraph{24 + vector like complex representations + real scalar $\phi$}~\\
Now we combine the above scenarios and add $n_{5+\bar{5}}=n_{10+\overline{10}}=n_{24}=1$ vector like fermions. This modifies the beta functions by $\Delta b_i = -18/3$ and adds $18$ zero-modes. We can again close all zero-modes with Yukawa couplings, such that we obtain for $SU(3)$
\be
\Lambda_{\rm SI}^4 \simeq 2 C_3 \left( \frac{8\pi^2}{g^2(M_{\rm GUT})}\right)^6 \left(\prod_i \frac{y_i}{4\pi}\right) \frac{y_{24}^3 \tilde{y}_{24}^5}{(4\pi)^{8}}\frac{y_5 y_{10}^3}{(4\pi)^4}\left(\frac{\Lambda_{SU(3)}}{M_{\rm GUT}}\right)^{b^{SU(3)}-4} \Lambda_{SU(3)}^4\,,
\ee
with $b^{SU(3)} = -1$. The $SU(2)_L$ contribution is given by
\be
\Lambda_{\rm SI}^4 \simeq 2 C_2 \left( \frac{8\pi^2}{g_w^2(M_{\rm GUT})}\right)^4 \left(\frac{c_{qqq\ell}}{16\pi^2}\right)^3 \frac{\tilde{g}_u \tilde{g}_d}{(4\pi)^2} \frac{y_{24} \tilde{y}_{24}^5}{(4\pi)^6}\frac{y_5 y_{10}^3}{(4\pi)^4} \left(\frac{\Lambda_{SU(2)}}{M_{\rm GUT}}\right)^{b^{SU(2)}-4} \Lambda_{SU(2)}^4\,,
\ee
where $b^{SU(2)} = -29/6$.
\paragraph{Superpartners at $10^6$~GeV}~\\
Including the scalar superpartners at a scale of $10^6$~GeV does not affect the form of the instanton contributions. It simply modifies the beta functions by $\Delta b^{SU(2)} = -13/6$ and $\Delta b^{SU(3)} = -2$.

\bibliographystyle{JHEP.bst}
\bibliography{draft}

\end{document}